\documentclass[12pt,a4]{article} 
\newcommand{\vs}{\vspace{-0.25cm}}
\usepackage{amsmath,graphicx}
\begin{document} 
\begin{center}
 {\Large{\bf Density-dependent NN-interaction from subsubleading chiral
3N-force: Intermediate-range contributions}\footnote{This work 
has been supported in part by DFG and NSFC (CRC110).}  }  

\medskip

 N. Kaiser \\
\medskip
{\small  Physik-Department T39, Technische Universit\"{a}t M\"{u}nchen,
   D-85747 Garching, Germany }
\end{center}
\medskip
\begin{abstract}
From the subsubleading chiral three-nucleon force [intermediate-range contributions, published in Phys. Rev.  C\,87, 054007 (2013)] a density-dependent NN-interaction $V_\text{med}$ is derived in isospin-symmetric nuclear matter. Following the division of the pertinent 3N-diagrams into two-pion-one-pion exchange topology and ring topology, one evaluates for these all selfclosings and concatenations of nucleon-lines to an in-medium loop. In the case of the $2\pi 1\pi$-exchange topology, the momentum- and $k_f$-dependent potentials associated with the isospin-operators ($1$ and $\vec\tau_1 \!\cdot\! \vec\tau_2$) and five independent spin-structures require at most one numerical integration. For the more challenging (concatenations of the) ring diagrams proportional to $c_{1,2,3,4}$, one ends up with regularized double-integrals $\int_0^\lambda  dr\,r \int_0^{\pi/2} d\psi$ from which the $\lambda^2$-divergence has been subtracted and the logarithmic piece $\sim \ln (m_\pi/\lambda)$ is isolated. The derived semi-analytical results are most helpful to implement the subsubleading chiral 3N-forces into nuclear many-body calculations.      
\end{abstract}

\section{Introduction and summary}
Three-nucleon forces are an indispensable ingredient in accurate few-nucleon and
nuclear structure calculations. Nowadays, chiral effective field theory is the
appropriate tool to construct systematically the nuclear interactions in harmony with the symmetries of QCD. Three-nucleon forces appear first at N$^2$LO, where they consist of a zero-range contact-term ($\sim c_E$), a mid-range $1\pi$-exchange component ($\sim c_D$) and a long-range $2\pi$-exchange component ($\sim c_{1,3,4}$). The complete calculation of the chiral 3N-forces to subleading order N$^3$LO \cite{3Nlong,3Nshort} and even to subsubleading order N$^4$LO \cite{twopi4,midrange4} has been achieved during the past decade by the Bochum-Bonn group. At present the focus lies on constructing 3N-forces in chiral effective field theory with explicit $\Delta(1232)$-isobars, for which the longe-range $2\pi$-exchange component  has been derived recently in ref.\,\cite{twopidelta} at order N$^3$LO.

However, for the variety of existing many-body methods, that are commonly employed in calculations of nuclear matter or medium mass and heavy nuclei, it is technically very challenging to include the chiral three-nucleon forces directly. An alternative and approximate approach is to use instead a density-dependent
two-nucleon interaction $V_\text{med}$ that originates from the underlying 3N-force. When restricting to on-shell scattering of two nucleons in isospin-symmetric spin-saturated nuclear matter, the resulting in-medium NN-potential  $V_\text{med}$ has the same isospin- and spin-structure as the free NN-potential.
The analytical expressions for $V_\text{med}$ from the leading chiral 3N-force at N$^2$LO (involving the parameters $c_{1,3,4}$, $c_D$ and $c_E$) have been presented in ref.\,\cite{holt} and these have found many applications (e.g. to thermodynamics of nuclear matter) in recent years \cite{corbinian1,corbinian2,samma,normalorder,achim1,achim2,achim3,carbone1,carbone2}. But in order to perform nuclear many-body calculations that are consistent with their input at the two-body level, one needs also $V_\text{med}$ derived from the subleading chiral 3N-forces at order N$^3$LO. In two recent works this task has been completed for the short-range terms and relativistic $1/M$-corrections in ref.\,\cite {vmedshort}, and for the long-range terms in ref.\,\cite {vmedlong}. In the latter case one is dealing with 3N-diagrams which were divided in ref.\,\cite{3Nlong} into classes of $2\pi$-exchange topology,  $2\pi1\pi$-exchange topology, and ring topology. For these topologies the selfclosings of a nucleon-line and the concatenations of any two nucleon-lines to an in-medium loop had to be worked out to together with the summation/integration over the filled Fermi-sea of density $\rho = 2k_f^3/3\pi^2$. The momentum- and $k_f$-dependent potentials associated with the isospin operators ($1$ and $ \vec\tau_1\!\cdot\!\vec\tau_2$) and five independent spin-structures ($1, \vec\sigma_1\!\cdot\!\vec\sigma_2, \vec\sigma_1\!\cdot\!\vec q\, \vec\sigma_2\!\cdot\! \vec q, $ $  i(\vec\sigma_1\!+\!\vec\sigma_2)\!\cdot\!(\vec q\!\times\!\vec p\,), $ $ \vec\sigma_1\!\cdot\! \vec p\, \vec\sigma_2\!\cdot\!\vec p +\vec \sigma_1\!\cdot\! \vec p\,' \vec \sigma_2\!\cdot\!\vec p\,')$ could
all be expressed in terms of functions, which were either given in closed analytical form or required at most one numerical integration. In order to obtain for the (non-factorizable) 3N-ring diagrams such an expedient form it was crucial to invert the order the original loop-integration and the added Fermi-sphere integral. Moreover, the method of dimensional regularization, as it was implicitly used in ref.\,\cite{3Nlong}, could be recovered by subtracting asymptotic constants from the integrands of $\int_0^\infty \!dl$. 

\begin{figure}[h]
\centering
\includegraphics[width=0.8\textwidth]{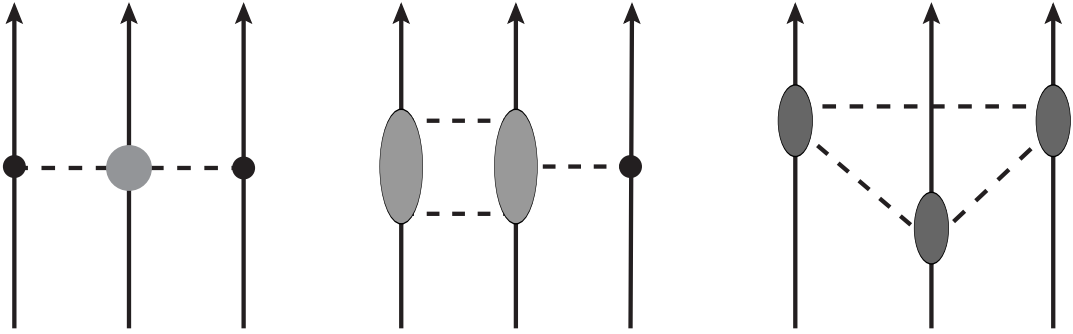}
\caption{$2\pi$-exchange topology, $2\pi1\pi$-exchange topology and ring topology which comprise the long- and  intermediate-range chiral 3N-forces at subsubleading order N$^4$LO.} \end{figure}

The pupose of the present paper is to extend the calculation of the in-medium NN-potential  $V_\text{med}$ to the subsubleading chiral 3N-forces at order N$^4$LO. The long-range $2\pi$-exchange component, symbolized by the left diagram in Fig.\,1, has already been treated in section\,4 of ref.\,\cite{vmedlong}  through approporiate contributions to the two structure functions $\tilde g_+(q_2)$ and  $\tilde h_-(q_2)$. As indicated by the notation, these structure functions are equal to $f_\pi^2$ times the isoscalar non-spin-flip and isovector  spin-flip $\pi N$-scattering amplitudes at zero pion-energy $\omega=0$ and squared momentum-transfer $t = -q_2^2$.  The present paper is organized as follows. We start in section 2 with the computation of $V_\text{med}$ from the intermediate-range $2\pi1\pi$-exchange component, symbolized by the middle diagram in Fig.\,1. In comparison to section\,3 of ref.\,\cite{vmedlong} one encounters at N$^4$LO a richer spin- and momentum-dependence for this part of the chiral 3N-force, and 12 instead of 8 functions $f_j(q_1)$ are needed to represent all diagrams belonging to this topology. The contributions to  $V_\text{med}$ as they arise from selfclosing, vertex-correction by $1\pi$-exchange, vertex-correction by $2\pi$-exchange, and double-exchange are given by semi-analytical expressions that comply with this extended structure. Note that 
ref.\,\cite{midrange4} has concluded from a study of the 3N-potential in coordinate space at the equilateral triangle configuration, that the N$^4$LO corrections to the intermediate-range topologies are numerically large and dominate in most cases over the nominally leading  N$^3$LO terms. This feature could be traced back to the large coefficients $c_{2,3,4}$, which reflect the importance of the $\Delta(1232)$-isobar coupled to the $\pi N$-system. At N$^4$LO the 3N-diagrams belonging to the ring topology, symbolized by the right diagram in Fig.\,1, fall into three classes according to their scaling with $g_A^2$. Section\,3 is devoted to the simpliest ring interaction proportional to $g_A^0 c_{1,2,3,4}$ and the contributions to $V_\text{med}$ from selfclosings and concatenations are given in three subsections. After angular integration the remaining double-integral $\int\!dl_0dl$ is treated in polar coordinates and regularized by a (euclidean) cutoff $\lambda$. In this form the $\lambda^2$-divergence can be easily subtracted and the subsequent  logarithmic piece $\sim \ln(m_\pi/\lambda)$ is isolated. A good check is provided by the fact that the total $\lambda^2k_f^3$-divergence is of isoscalar central type and thus can be absorbed on the 3N short-distance parameter $c_E$. In section\,4 the analogous calculations are carried out for the more involved ring interaction proportional to $g_A^2 c_{1,2,3,4}$. Finally, one considers in section\,5 the ring interaction proportional to $g_A^4 c_{1,2,3,4}$, which consists of a large number of terms with different isospin-, spin- and momentum-dependence. At that point we elaborate also a bit on euclidean loop-integrals over four or three pion-propagators. The selfclosing contributions to $V_\text{med}$ are given in closed analytical form in subsection\,5.1 and one observes that these central, spin-spin and tensor potentials linear in density $\rho$ depend either on $c_2+c_3$ or on $c_1$ and $c_3$. Concerning the contributions to $V_\text{med}$ from concatenations, we present in subsection\,5.2 the pertinent expressions only for three selected (yet simple) terms from the ring interaction $\sim g_A^4 c_{1,2,3,4}$. These give rise either to an isovector spin-orbit potential,  or to isoscalar and isovector central and spin-spin potentials.  A complete list of the lengthy formulas for the remaining contributions to $V_\text{med}$ from the concatenations of the 3N-ring interaction $\sim g_A^4 c_{1,2,3,4}$ can be obtained from the author upon request.

In summary, after eventual partial-wave projection the presented results for $V_\text{med}$ are suitable for an approximate implementention of the subsubleading  chiral 3N-forces of intermediate range into nuclear many-body calculations.

\section{Two-pion-one-pion exchange topology}
The $2\pi1\pi$-exchange 3N-interaction arises from a large set of loop-diagrams, and according to eq.(3.1) in ref.\,\cite{midrange4} it  can be written in the general form:
\begin{eqnarray}
V_\text{3N}&=&{g_A^4 \over 256\pi f_\pi^6} { \vec\sigma_3\!\cdot\! \vec q_3 \over m_\pi^2+q_3^2}\Big\{ \vec\tau_1 \!\cdot\! \vec\tau_3\Big[\vec\sigma_2\!\cdot\! \vec q_1\,\vec q_1\!\cdot\! \vec q_3\, f_1(q_1) + \vec\sigma_2\!\cdot\! \vec q_1\,f_2(q_1)+ \vec\sigma_2\!\cdot\! \vec q_3\,f_3(q_1)\Big] \nonumber\\ &&   +\vec\tau_2\!\cdot\! \vec\tau_3\Big[\vec\sigma_1\!\cdot\! \vec q_1\,\vec q_1\!\cdot\! \vec q_3\, f_4(q_1) + \vec\sigma_1\!\cdot\! \vec q_3\,f_5(q_1)+\vec\sigma_2\!\cdot\! \vec q_1\,\vec q_1\!\cdot \!\vec q_3\, f_6(q_1)\nonumber\\ &&+\vec\sigma_2\!\cdot\! \vec q_1\, f_7(q_1)+ \vec\sigma_2\!\cdot\! \vec q_3\,\vec q_1\!\cdot \!\vec q_3\,f_8(q_1)+ \vec\sigma_2\!\cdot\! \vec q_3\,f_9(q_1)\Big] \nonumber\\ &&  +(\vec\tau_1\!\times \!\vec\tau_2)\!\cdot\! \vec\tau_3\, \Big[(\vec\sigma_1\!\times\!\vec\sigma_2)\!\cdot\!\vec q_1\Big(\vec q_1\!\cdot \!\vec q_3\, f_{10}(q_1)+f_{11}(q_1)\Big) + \vec \sigma_1\!\cdot\!(\vec q_1\!\times \!\vec q_3)\,\vec \sigma_2\!\cdot\!\vec q_1\, f_{12}(q_1)\Big] \Big\}\,,\end{eqnarray}
where $\vec q_j$ denotes the momentum-transfer at nucleon $j\in\{1,2,3\}$, and  $\vec q_1+\vec q_2+\vec q_3=0 $ holds due to momentum-conservation. Since a common prefactor $g_A^4/(256\pi f_\pi^6)$ has been pulled out in eq.(1), the contributions  to the reduced functions $f_j(s)$ at N$^3$LO read according to eq.(3.2) in ref.\,\cite{midrange4}:    
\begin{eqnarray}&& f_1(s) = {m_\pi\over s^2}(1-2 g_A^2)-{g_A^2m_\pi\over 4m_\pi^2 + s^2}  +\Big[1+g_A^2 +{4 m_\pi^2 \over s^2}(2 g_A^2-1)\Big] A(s) \,,
\\ && f_2(s)= f_7(s)=(4m_\pi^2+2s^2)A(s) \,,
\\ && f_3(s)= \big[4(1-2g_A^2)m_\pi^2+(1-3g_A^2)s^2\big]A(s) \,, \\  &&  f_5(s)=-s^2 f_4(s)=2 g_A^2 s^2 A(s)\,, \\ && f_{11}(s)= -\Big(2m_\pi^2+{s^2\over 2}\Big)A(s) \,,\\
&& f_{6,8,9,10,12}(s)=0\,, \end{eqnarray}
with the heavy-baryon loop-function 
\begin{equation}A(s) = {1\over 2s}\arctan{s \over 2m_\pi}\,. \end{equation}
Likewise, one extracts from eq.(3.3) in ref.\,\cite{midrange4} the following contributions to  the reduced functions $f_j(s)$ at N$^4$LO: 
\begin{eqnarray}&& f_1(s) = {16 c_4 \over 3\pi} \bigg\{(4-g_A^{-2}){m_\pi^2\over s^2}
+\bigg[ (g_A^{-2}-4){m_\pi^2\over s^2}+{1-g_A^{-2}\over 2}-{3m_\pi^2\over 4m_\pi^2 + s^2}\bigg] L(s)\bigg\}\,, \\ 
&& f_3(s) = {16 c_4 \over 3\pi} \bigg[ (g_A^{-2}-1)m_\pi^2+(g_A^{-2}-4)s^2-
{12m_\pi^4\over 4m_\pi^2 + s^2}\bigg] L(s)\,, \\  
&&  f_5(s)=-s^2 f_4(s)={16 c_4 \over \pi} s^2 L(s)\,, \end{eqnarray}
\newpage
\begin{eqnarray}
&&  f_6(s)= {8 \over 3\pi} \bigg\{(6c_1+c_2-3c_3){m_\pi^2\over s^2}+\bigg[(3c_3-6c_1-c_2)
{m_\pi^2\over s^2}+{c_2\over 2}+{3(2c_1+c_3) m_\pi^2 \over 4m_\pi^2+s^2}\bigg]L(s)\bigg\} \,,\\ 
&& f_7(s)= {8 \over \pi g_A^2} \bigg\{\bigg[2\Big(c_3+{c_2 \over 3}-2c_1\Big)
m_\pi^2+\Big({c_2\over 6}+c_3\Big)s^2\bigg]L(s)+\Big( (8\pi f_\pi)^2 \bar e_{14} -{5c_2 \over 18} -c_3\Big){s^2\over 2}\bigg\} \,, \\ 
&& f_9(s)= {8 \over \pi }\bigg[(8c_1-c_2-4c_3)m_\pi^2-(3c_2+13c_3){s^2\over 4}-{4(2c_1+c_3)m_\pi^4 \over 4m_\pi^2+s^2} \bigg]L(s)\,, \\ 
&& f_{10}(s) = f_{12}(s) = {4c_4 \over \pi} L(s)\,, \\
&& f_{2,8,11}(s)=0\,. \end{eqnarray}
with the frequently occurring logarithmic loop-function
 \begin{equation}L(s) = {\sqrt{4m_\pi^2+s^2}\over s}\ln{s+ \sqrt{4m_\pi^2+s^2}\over 2m_\pi}\,. \end{equation}
Note that we have supplied in eq.(13) through the last term proportional to $s^2/2$ that particular polynomial piece\footnote{This important additional information was  provided by H. Krebs. The value of the low-energy constant $\bar e_{14}$ as extracted from $\pi N$-scattering is $\bar e_{14}= 1.52$\,GeV$^{-3}$\,\cite{twopi4} or $\bar e_{14}= 1.18$\,GeV$^{-3}$\,\cite{hofer}.} which cannot be absorbed on the short-distance parameters $c_D$ and $c_E$.  
One notices from eqs.(7,16) that there is yet no contribution to $f_8(s)$, but the corresponding structure 
$\vec\sigma_2\!\cdot\! \vec q_3\,\vec q_1\!\cdot \!\vec q_3$ will arise once explicit $\Delta(1232)$-isobars are considered in the derivation of the chiral $2\pi1\pi$-exchange 3N-interaction.  
\subsection{Contributions to in-medium NN-potential}
Now we turn to the contributions of the $2\pi1\pi$-exchange 3N-interaction $V_\text{3N}$ written in eq.(1) to the in-medium NN-potential $V_\text{med}$. Only the selfclosing of nucleon line\,1 gives a non-vanishing spin-isospin trace, and after relabeling $3\to1$ one obtains the contribution 
\begin{equation}V^{(0)}_\text{med} = {g^4_A k^3_f f _9(0) \over 3(4\pi f^2_\pi)^3}
{\vec\tau_1\!\cdot\!\vec\tau_2\over m^2_\pi+q^2} \,\vec\sigma_1\!\cdot\!\vec
q\, \vec\sigma_2\!\cdot\!\vec q=   {g^4_A m_\pi^2 k^3_f\over 24
\pi^4 f_\pi^6} {\vec\tau_1\!\cdot\!\vec\tau_2\over m^2_\pi+q^2} \,\vec\sigma_1\!\cdot\!\vec
q\, \vec\sigma_2\!\cdot\!\vec q\,(6c_1-c_2-5c_3)\,,\end{equation}
which is of the form: $1\pi$-exchange NN-interaction times a factor linear in density $\rho=2k_f^3/3\pi^2$.
The last expression in eq.(18) comes from evaluating $f_9(s)$ in eq.(14) at $s=0$. In all forthcoming formulas for $V_\text{med}$ we denote by $\vec q = \vec p\,' - \vec p$ the momentum-transfer for the on-shell scattering process $N_1(\vec p\,)+N_2(-\vec p\,)\to N_1(\vec p\,')+N_2(-\vec p\,')$ in the nuclear matter rest-frame. On the other hand the vertex corrections by $1\pi$-exchange, apparent in eq.(1) through the factor $\vec\sigma_3\!\cdot\vec q_3/(m_\pi^2+q_3^2)$, produce the contribution
\begin{eqnarray}V^{(1)}_\text{med} &=&  {g_A^4 \over (8\pi f_\pi^2)^3} \bigg\{ 
\Big(2m_\pi^2\Gamma_0-{4k_f^3\over 3}\Big)\big[\vec\tau_1\!\cdot\!\vec\tau_2 f_3(q)+3f_9(q) \big] - \Big(2\Gamma_2 +{q^2 \over 2} \widetilde\Gamma_3\Big)  q^2 \big[ \vec\tau_1\!\cdot\!\vec\tau_2 f_1(q)+3 f_6(q)\big] \nonumber\\ && + \widetilde\Gamma_1 q^2 \big[\vec \tau_1\!\cdot\!\vec\tau_2 f_2(q)+3 f_7(q) \big]+(2k_f^3-3m_\pi^2 \widetilde\Gamma_1)q^2 f_8(q) -3 (\vec\sigma_1\!\cdot\!\vec p\, \vec\sigma_2\!\cdot\!\vec p+\vec\sigma_1\!\cdot\!\vec p\,'\, \vec\sigma_2\!\cdot\!\vec p\,')\widetilde \Gamma_3 f_5(q) 
\nonumber\\ && +2\vec\sigma_1\!\cdot\!\vec\sigma_2\Big[-3 \Gamma_2 f_5(q)+\vec\tau_1\!\cdot\!\vec\tau_2 \,q^2\Big(2 \Gamma_2\big[f_{10}(q)+f_{12}(q)\big]-\widetilde\Gamma_1 f_{11}(q)+{q^2 \over 2}\widetilde\Gamma_3 f_{10}(q) \Big)\Big] \nonumber\\ && 
+ 2\vec\sigma_1\!\cdot\!\vec q\, \vec\sigma_2\!\cdot\!\vec q\, \Big[\vec\tau_1\!\cdot\!\vec\tau_2 \Big(\widetilde\Gamma_1 f_{11}(q)-2 \Gamma_2\big[f_{10}(q)+f_{12}(q)\big]
-{q^2 \over 2} \widetilde\Gamma_3 f_{10}(q)\Big) 
-3\Big(\Gamma_2 +{q^2 \over 4} \widetilde\Gamma_3\Big) f_4(q) \Big] \nonumber\\ && 
+4 \vec \tau_1\!\cdot\!\vec\tau_2  \, \vec\sigma_1\!\cdot\!(\vec q\!\times\!\vec p\,) \, \vec\sigma_2\!\cdot\!(\vec q\!\times\!\vec p\,) \widetilde\Gamma_3 f_{12}(q)\
+i( \vec\sigma_1\!+\!\vec\sigma_2)\!\cdot \! (\vec q\!\times \! \vec p\,) \Big[ {3q^2 \over 2}\widetilde\Gamma_3 f_6(q) -3 \widetilde\Gamma_1 f_7(q)\nonumber\\ && +\vec\tau_1\!\cdot\!\vec\tau_2 \Big(\widetilde\Gamma_1\big[2f_{11}(q)-f_2(q)\big] +{q^2\over 2}\widetilde\Gamma_3 \big[f_1(q)-2f_{10}(q)+2f_{12}(q)\big] \Big) \Big] \bigg\}\,, \end{eqnarray}
with the $(p,k_f)$-dependent functions $\Gamma_0,\, \widetilde\Gamma_1= \Gamma_0 +\Gamma_1,\,\Gamma_2$ and  $\widetilde\Gamma_3= \Gamma_0 +2\Gamma_1+\Gamma_3$ defined in the appendix of ref.\,\cite{vmedshort}. Moreover, the vertex corrections by $2\pi$-exchange, represented by the expression in curly brackets of eq.(1), can be summarized as the  $1\pi$-exchange NN-interaction times a $(p,q,k_f)$-dependent factor 
\begin{equation}V^{(2)}_\text{med} = {g^4_A \over (8\pi f^2_\pi)^3}
{\vec\tau_1\!\cdot\!\vec\tau_2\over m^2_\pi+q^2} \,\vec\sigma_1\!\cdot\!\vec
q\, \vec\sigma_2\!\cdot\!\vec q\,\big[ S_1(p,k_f) +q^2 S_2(p,k_f)\big]\,.\end{equation}
The two auxiliary functions $S_{1,2}(p,k_f)$ are computed as integrals over $f_j(s)$ in the following way:
\begin{eqnarray}S_1(p,k_f) &=& \int_{p-k_f}^{p+k_f}\!\! ds\, {s \over p}\big[k_f^2-(p-s)^2\big] \bigg\{ 2s^2 f_{12}(s)-f_3(s)-f_5(s)-f_9(s) \nonumber \\ && +{1\over 8p^2}\big[(p+s)^2-k_f^2\big]\big[f_2(s)+ f_7(s)-4f_{11}(s)\big] +{1\over 24p^2}\big[k_f^2-(p-s)^2\big]\nonumber\\ && \times (s^2+4s p +p^2-k_f^2)\big[4f_{10}(s)-2 f_{12}(s)-f_1(s)-f_4(s)-f_6(s)\big] \bigg\} \,, \end{eqnarray}
\begin{eqnarray}S_2(p,k_f) &=& \int_{p-k_f}^{p+k_f}\!\! ds\, {s \over 8p^3}\big[k_f^2-(p-s)^2\big]\big[(p+s)^2-k_f^2\big]\bigg\{f_8(s) \nonumber\\ && +{1\over 4p^2}(s^2 +p^2-k_f^2)\big[4f_{10}(s)-2 f_{12}(s)-f_1(s)-f_4(s)-f_6(s)\big] \bigg\} \,. \end{eqnarray}
Finally, there is the contribution $V^{(3)}_\text{med}$ from the double-exchange. We separate it into an isoscalar part:
\begin{eqnarray}V^{(3)}_\text{med}&=&{3g_A^4\over (8\pi f_\pi^2)^3}\bigg[\vec\sigma_1\!\cdot\!\vec \sigma_2\Big(2I_{2,2}-2I_{3,2}-H_{1,2}-\tilde I_{1,2}\Big)+ \vec\sigma_1\!\cdot\!\vec q\, \vec \sigma_2\!\cdot\!\vec q \,\Big({H_{1,1}+\tilde I_{1,4}\over 2}-I_{2,4}-I_{3,5}\Big) \nonumber\\ && +(\vec\sigma_1\!\cdot\!\vec p\, \vec\sigma_2\!\cdot\!\vec p+\vec\sigma_1\!\cdot\!\vec p\,'\, \vec\sigma_2\!\cdot\!\vec p\,')\Big(I_{2,3}-I_{3,3}-{H_{1,3}+\tilde I_{1,3}\over 2}\Big)\bigg]\,, \end{eqnarray}
and an isovector part:
\begin{eqnarray}V^{(3)}_\text{med}&=&{g_A^4\vec\tau_1\!\cdot\!\vec \tau_2\over(8\pi 
f_\pi^2)^3}\bigg\{2m_\pi^2 I_{5,0}-2H_{5,0}+{q^2 \over 2}(H_{4,1}\!+\!\tilde I_{4,4})-p^2(H_{4,3}\!+\!\tilde I_{4,3})-3H_{4,2}-3 \tilde I_{4,2} \nonumber\\ && +i(\vec\sigma_1\!+\!\vec\sigma_2)\!\cdot\!(\vec p\!\times\!\vec q\,)\Big[H_{10,1}+\tilde I_{10,1}-{1\over 2}(H_{4,1}+\tilde I_{4,1})- 2I_{11,1}\Big] \nonumber\\ && +\vec \sigma_1\!\cdot\!\vec \sigma_2\Big[2 I_{7,2}-H_{6,2}-\tilde I_{6,2}+ H_{8,2} +\tilde I_{8,2}-2I_{9,2}
+4H_{10,2}+4\tilde I_{10,2}-8 I_{11,2}+2H_{12,2} \nonumber\\ && -2\tilde I_{12,2}-4m_\pi^2 I_{12,2}+2p^2\big(H_{10,3}+\tilde I_{10,3}-2 I_{11,3}\big) -q^2\big( H_{10,1}+\tilde I_{10,4}-2I_{11,4}\big)\Big]\nonumber\\ && +\vec\sigma_1\!\cdot\!\vec q\, \vec \sigma_2\!\cdot\!\vec q \,\Big[{H_{6,1}+\tilde I_{6,4}+\tilde I_{8,5}\over 2} -I_{7,4}+H_{8,0}-H_{8,1}-I_{9,5}+H_{10,1} +\tilde I_{10,4} \nonumber\\ && -2I_{11,4}+H_{12,1}+\tilde I_{12,4}+2m_\pi^2\big(2I_{12,4}+ I_{12,5}-2I_{12,0}\big)\Big]   +(\vec\sigma_1\!\cdot\!\vec p\, \vec\sigma_2\!\cdot\!\vec p+\vec\sigma_1\!\cdot\!\vec p\,'\, \vec\sigma_2\!\cdot\!\vec p\,')\nonumber\\ && \times \Big[{H_{8,3}\!-\!H_{6,3}\!+\!\tilde I_{8,3}\!-\!\tilde I_{6,3} \over 2} +I_{7,3}-I_{9,3}-H_{10,3}-\tilde I_{10,3}+2 I_{11,3}+H_{12,3}- \tilde I_{12,3}-2m_\pi^2 I_{12,3}\Big]\bigg\}\,. \nonumber\\ \end{eqnarray}
The double-indexed functions $H_{j,\nu}(p)$ are defined by:
\begin{equation} H_{j,0}(p) = {1\over 2p}\int_{p-k_f}^{p+k_f}\!\!ds\, s f_j(s)\big[k_f^2-(p-s)^2\big]\,, \end{equation} 
\begin{equation} H_{j,1}(p) = {1\over 8p^3}\int_{p-k_f}^{p+k_f}\!\!ds\, s f_j(s)\big[k_f^2-(p-s)^2\big]\big[(p+s)^2-k_f^2\big]\,, 
\end{equation} 
\begin{equation} H_{j,2}(p) = {1\over 48p^3}\int_{p-k_f}^{p+k_f}\!\!ds\, s f_j(s)\big[k_f^2-(p-s)^2\big]^2(s^2+4s p +p^2-k_f^2)\,, 
\end{equation} 
\begin{equation} H_{j,3}(p) = {1\over 16p^5}\int_{p-k_f}^{p+k_f}\!\!ds\, s f_j(s)\big[k_f^2-(p-s)^2\big]\big[(p+s)^2-k_f^2\big](p^2+s^2-k_f^2)\,. \end{equation} 
The other double-indexed functions  $I_{j,\nu}(p,q)$  are defined by: 
\begin{equation} I_{j,0}(p,q) = {1\over 2q}\int_{p-k_f}^{p+k_f}\!\!ds\, s f_j(s)
\ln{q X +2\sqrt{W} \over (2p+q)[m_\pi^2+(s-q)^2]}\,, \end{equation} \begin{equation} I_{j,1}(p,q) = {1\over 4p^2-q^2}\int_{p-k_f}^{p+k_f}\!\!ds\, s f_j(s)\bigg[ {p(s^2+m_\pi^2)-\sqrt{W} \over q^2}+{p^2+k_f^2-s^2 \over 2p}\bigg]\,, \end{equation} 
\begin{eqnarray} I_{j,2}(p,q) &\!\!\!\!\!\!\!\!=\!\!\!\!\!\!\!\!& {1\over 8q^2}\int_{p-k_f}^{p+k_f}\!\!ds\, s f_j(s)\bigg\{s(m_\pi^2+s^2+q^2)-p\Big(m_\pi^2+s^2+{3q^2 \over 4}\Big)\nonumber\\ && -{1\over 2q}\big[m_\pi^2+(s+q)^2\big] \big[m_\pi^2+(s-q)^2\big] \ln{q X +2\sqrt{W} \over (2p+q)[m_\pi^2+(s-q)^2]}  \nonumber\\ && - {X \sqrt{W} \over 4p^2-q^2} -{(k_f^2-s^2)^2 \over p} +{p\over  4p^2-q^2}\Big(m_\pi^2+2k_f^2 -s^2+{q^2\over 2}\Big)^2  \bigg\}\,, \end{eqnarray}
\begin{eqnarray} I_{j,3}(p,q) &\!\!\!\!\!\!\!\!=\!\!\!\!\!\!\!\!& {1\over (4p^2-q^2)^2}\int_{p-k_f}^{p+k_f}\!\!ds\, s f_j(s)\bigg\{ {X \sqrt{W} \over q^2} +{q^2 \over 8p^3}(k_f^2-s^2)^2 -{3pq^2\over 8}\nonumber\\ &&+{p \over q^2}(s^2+m_\pi^2)(2p^2+s^2-2k_f^2-m_\pi^2) +{p \over 2}(2k_f^2-3m_\pi^2+p^2-s^2) \nonumber\\ && +{1\over 4p}\Big[s^2(2m_\pi^2+q^2-4s^2)+k_f^2(10s^2-2m_\pi^2-3q^2)-6k_f^4\Big]  \bigg\}\,, \end{eqnarray}
\begin{eqnarray} I_{j,4}(p,q) &\!\!\!\!\!\!\!\!=\!\!\!\!\!\!\!\!& {1\over 4q^4}\int_{p-k_f}^{p+k_f}\!\!ds\, s f_j(s)\bigg\{ {X\over (4p^2-q^2)^2}\Big[\sqrt{W}(3q^2-4p^2)- 8p^3X\Big]\nonumber\\ && + \bigg[ {q^3\over 2}+q(s^2-m_\pi^2) -{3\over 2q}(s^2+m_\pi^2)^2\bigg] \ln{q X +2\sqrt{W} \over (2p+q)[m_\pi^2+(s-q)^2]} \nonumber\\ && +{p \over 4p^2-q^2}\Big[16k_f^4+8k_f^2(2m_\pi^2+q^2-2s^2 )+3m_\pi^4+3s^4+2q^2(m_\pi^2-s^2)-10s^2m_\pi^2\Big] \nonumber\\ && +s(3s^2-q^2+3m_\pi^2)+2p^3-p(4k_f^2+4m_\pi^2+q^2) +{2\over p}(k_f^2+q^2-s^2)(s^2-k_f^2)  \bigg\}\,,  \end{eqnarray}
\begin{equation} I_{j,5}(p,q) =-I_{j,4}(p,q)+{1\over 2q^2}\int_{p-k_f}^{p+k_f}\!\!ds\, s f_j(s)\bigg[{q^2\!-\!s^2\!-\!m_\pi^2\over q} \ln{q X +2\sqrt{W} \over (2p+q)[m_\pi^2+(s-q)^2]}+{k_f^2-(p-s)^2\over p}\bigg]\,. \end{equation} 
with the auxiliary polynomials
\begin{eqnarray} && X = m_\pi^2 +2(k_f^2-p^2)+q^2-s^2\,, \nonumber \\
&& W = k_f^2 q^4+p^2(m_\pi^2+s^2)^2 +q^2\big[ (k_f^2-p^2)^2+m_\pi^2(k_f^2+p^2)
   -s^2(k_f^2+p^2+m_\pi^2)\big]\,.  \end{eqnarray}
Furthermore, the functions  $\tilde I_{j,\nu}(p,q)$ with $j = 1,4,6,8,10,12$ appearing in eqs.(23,24) are computed analogously by replacing in the integrand $f_j(s)$ by  $\tilde f_j(s)= (s^2-m_\pi^2-q^2)f_j(s) $. The decomposition into $H_{j,\nu}$ and  $I_{j,\nu}$ is obtained by canceling momentum-factors against a pion-propagator, while  $\tilde I_{j,\nu}$  takes care of $s^2$-dependent remainder terms.

\section{Ring interaction proportional to $g_A^0$}
Next, we turn to the 3N-ring interaction at N$^4$LO, which consists of three pieces with different  dependence on the axial-vector coupling constant: $g_A^{2n},\, n=0,1,2$. The $g_A^0$-part  can be obtained directly from the well-known Feynman rules for the $\pi\pi NN$ Tomozawa-Weinberg vertex and the second-order  $\pi\pi NN$-contact vertex proportional to $c_{1,2,3,4}$. Altogether, the 3N-ring interaction proportional to $g_A^0 c_{1,2,3,4}$ is given by a euclidean loop-integral of the form 
\begin{eqnarray}V_\text{3N}&=&-{1\over f_\pi^6}\int_0^\infty\!\! dl_0\!\int\!{ d^3l_2\over (2\pi)^4} {l_0^2\over (\bar m^2+l_1^2)(\bar m^2+l_2^2)(\bar m^2+l_3^2)} \nonumber\\ && \times \Big\{\vec \tau_2\!\cdot\!\vec \tau_3\big[2c_1 m_\pi^2+(c_2+c_3)l_0^2 +c_3 \,\vec l_2\!\cdot\!\vec l_3\big]+{c_4 \over 4}\vec\tau_1\!\cdot\!(\vec \tau_2\!\times\!\vec \tau_3) \,\vec \sigma_1\!\cdot\!(\vec l_3\!\times\!\vec l_2)\Big\}\,, \end{eqnarray}
with $\bar m = \sqrt{m_\pi^2+l_0^2}$ and one has to set $\vec l_1= \vec l_2-\vec q_3$ and $\vec l_3= \vec l_2+\vec q_1$. Alternatively, one can take the (Fourier-transformed) coordinate-space potential in eq.(4.8) of ref.\,\cite{midrange4} and translate spatial gradients back to momentum factors $\vec l_{1,2,3}$. Note that the  4-dimensional loop integral in eq.(36) is quadratically divergent and therefore the 3N-ring interaction $V_\text{3N}$ requires a regularization (e.g. by
an ultraviolet cutoff) and a  renormalization (by absorbing cutoff-dependent
pieces on the 3N short-distance parameters $c_E$ and $E_{1,\dots,10}$
\cite{3Ncontact}).

\subsection{In-medium NN-potential from selfclosing of nucleon-lines}
Only the selfclosing of nucleon line\,1 gives a non-vanishing spin-isospin trace, and after relabeling $3\to1$ one recognizes an isovector central potential $\sim k_f^3 \vec \tau_1\!\cdot\!\vec \tau_2$. Evaluating the pertinent loop-integral in spherical coordinates $l_0 = r \cos\psi,  l_2 = r \sin\psi, \hat l_2\!\cdot\! \hat q = \cos\theta$ and introducing a cutoff $\lambda$  for the radial integral $\int_0^\lambda\!dr$, one gets the following  contribution to the in-medium NN-potential:
\begin{eqnarray}V_\text{med}^{(0)} &=&{ k_f^3 \,\vec \tau_1\!\cdot\!\vec \tau_2 \over 48 \pi^4 f_\pi^6} \bigg\{ \bigg[ \Big( 2c_1-{3c_2 \over 2}-3c_3\Big) m_\pi^2-\Big( {c_2 \over 4}+{c_3 \over 3}\Big) q^2 \bigg] \ln{m_\pi \over \lambda} +\Big( {3c_2 \over 2}-2c_1+{13 c_3\over 3}\Big) {m_\pi^2\over 4}\nonumber\\ && +\Big({13c_2 \over 8}+{11c_3\over 3}\Big) {q^2 \over 12} +\bigg[ \Big( 2c_1-c_2-{7c_3 \over 3}\Big) m_\pi^2 - \Big( {c_2 \over 4}+{c_3 \over 3}\Big) q^2\bigg] L(q)\bigg\}\,,  \end{eqnarray}
with the function $L(q)$  defined in eq.(17). Note that the power divergence proportional to $\lambda^2 k_f^3$ has been dropped in eq.(37), but it will be considered in the total balance   at the end of this section. Note also that the coefficient $c_2/4+c_3/3$ appears twice, such that the chiral limit $m_\pi \to 0$ of $V_\text{med}^{(0)}$ exists.
\subsection{In-medium potential from concatenations $N_3$ on $N_2$ and $N_2$ on $N_3$ }
\begin{table}[h!]
\begin{center}
\begin{tabular}{|c|c|c|c|c|c|c|}
    \hline
concat. & $N_3$ on $N_2$  & $N_2$ on $N_3$ &  $N_3$ on $N_1$  & $N_1$ on $N_3$ &  $N_1$ on $N_2$  & $N_2$ on $N_1$  \\  \hline 

$\vec l_1=$  & $\vec l_4+\vec l$  & $-\vec l_4-\vec l$  & $\vec l-\vec p\,'$ &  $\vec p-\vec l$  & $\vec l-\vec p$ &  $\vec p\,'-\vec l$ \\ 

$\vec l_2=$ & $\vec l-\vec p\,'$  &  $\vec p-\vec l$  & $\vec l_4+\vec l$ & $-\vec l_4-\vec l$ & $\vec l-\vec p\,'$ &  $\vec p-\vec l$\\ 

$\vec l_3=$  & $\vec l-\vec p$ &  $\vec p\,'-\vec l$ & $\vec l-\vec p$ & $\vec p\,'-\vec l$ & $\vec l_4+\vec l$ & $-\vec l_4-\vec l$  \\ 
  \hline
  \end{tabular}
\end{center}
{\it Tab.1: Assignment of pion momenta, where  $\vec l$ is unconstrained and 
$|\vec l_4|<k_f$ from a Fermi sphere.}
\end{table}
Next, one has to work out for $V_\text{3N}$ in eq.(36) the six possible concatenations of two nucleon-lines and their mirror graphs. The proper assignments of $\vec l_1,\, \vec l_2,\,\vec l_3$, with $\vec l$ the unconstrained loop-momentum and $\vec l_4$ from the interior  of a Fermi-sphere $|\vec l_4|<k_f$, are given for each concatenation in Table\,1.  The integral over the Fermi-sphere and the angular part of the loop-integral can always be solved analytically in terms of the following  functions:
\begin{equation} \bar\Gamma_0(l) = k_f-\bar m\bigg[  \arctan{k_f+l\over \bar m}+ \arctan{k_f-l\over \bar m}\bigg]+{\bar m^2+k_f^2-l^2\over 4 l} \ln{\bar m^2+(k_f+l)^2\over \bar m^2+(k_f-l)^2}\,,  \end{equation}
\begin{equation} \bar\Gamma_1(l) = {k_f\over 4l^2}(\bar m^2+k_f^2+l^2) -{1\over 16l^3}\big[\bar m^2+(k_f+l)^2\big]\big[\bar m^2+(k_f-l)^2\big] \ln{\bar m^2+(k_f+l)^2\over \bar m^2+(k_f-l)^2}\,,  \end{equation}
\begin{equation}\bar\Gamma_2(l) = {k_f^3\over 9}- {\bar m^2\over 3}\bar\Gamma_0(l)+{1\over 6}(k_f^2+\bar m^2-l^2) \bar\Gamma_1(l)  \,,  \end{equation}
\begin{equation}\bar\Gamma_3(l) = {k_f^3\over 3l^2}+{l^2-\bar m^2-k_f^2\over 2 l^2} \bar\Gamma_1(l)  \,,  \end{equation}
\begin{equation} \Lambda(l)={1\over 4p}\ln {\bar m^2+(l+p)^2 \over \bar m^2+(l-p)^2}\,,  \end{equation}
\begin{equation} \Omega(l)={1\over q \sqrt{B+q^2l^2}}\ln {q\, l +\sqrt{B+q^2l^2} \over \sqrt{B}}\,,  \end{equation} with the abbreviation $B= [\bar m^2+(l+p)^2][\bar m^2+(l-p)^2]$.
For remaining integration over $dl_0 dl$ one chooses polar coordinates $l_0 = r \cos\psi,\, l = r \sin\psi$ and sets a radial cutoff $\lambda$. By performing these calculational steps one obtains from the 
concatenations $N_3$ on $N_2$ and $N_2$ on $N_3$ an isoscalar central potential of the form
\begin{eqnarray} V_\text{med}^{(1)}&=& {3 \over 4\pi^5 f_\pi^6} \int_0^\lambda\!\! dr r \!\int_0^{\pi/2}\!\!\!d\psi \bigg\{ l_0^2 l \,\bar \Gamma_0(l)\Big[c_3 \Lambda(l)+ \Big( 2c_1 m_\pi^2+c_2 l_0^2 -{c_3\over 2}(2m_\pi^2+q^2) \Big)\Omega(l)\Big] \nonumber \\ &&  -{k_f^3\over 6}(c_3+c_2 \cos^2 \psi) \sin^22\psi\bigg\}\,. \end{eqnarray}
The purpose of the subtraction term in the second line is to remove a power divergence proportional to $\lambda^2 k_f^3$. After that the double-integral in eq.(44) has only a logarithmic dependence on the cutoff:
\begin{equation} {\pi k_f^3 \over 48 }\bigg[ c_2\Big( 3m_\pi^2+{3k_f^2 \over 10}+{p^2 \over 2}+{q^2 \over 4}\Big) -4c_1 m_\pi^2 +c_3\Big(6m_\pi^2+{2k_f^2\over 5}+{2p^2\over 3}+q^2\Big) \bigg] \ln {m_\pi \over \lambda} \,, \end{equation} 
and this detailed knowledge may be useful for numerical checks.  

The last $c_4$-term in eq.(36) produces in the same way a contribution to the isovector spin-orbit potential

\begin{equation} V_\text{med}^{(1)}= {c_4  \vec \tau_1\!\cdot\!\vec \tau_2\over 8\pi^5 f_\pi^6} i(\vec \sigma_1\!+\!\vec \sigma_2)\!\cdot\!(\vec q\!\times\!\vec p\,) \int_0^\lambda\!\! dr r \!\int_0^{\pi/2}\!\!
\!d\psi \, { l_0^2 l \,\bar \Gamma_0(l)\over 4p^2-q^2} \Big[ \Lambda(l) +\Big(p^2-l^2-\bar m^2-{q^2\over 2}\Big)  \Omega(l)\Big]  \,, \end{equation}
with a large-$\lambda$ behavior of the double-integral: $-(\pi k_f^3 /144) \ln(m_\pi/ \lambda)$.
\subsection{In-medium NN-potential from remaining  four concatenations}
The other four concatenations,  $N_3$ on $N_1$,  $N_1$ on $N_3$, $N_1$ on $N_2$, and  $N_2$ on $N_1$, applied to the $c_{1,2,3}$-term in eq.(36) give rise to an isovector central potential of the form
\begin{eqnarray} V_\text{med}^{(\text{cc})}&=& {\vec \tau_1\!\cdot\!\vec \tau_2 \over 4\pi^5 f_\pi^6} 
\int_0^\lambda\!\! dr r \!\int_0^{\pi/2}\!\!\! d\psi \bigg\{ l_0^2 l \Big\{ \big[ 4c_1 m_\pi^2+2(c_2+c_3) l_0^2\big] \bar \Gamma_0(l) \Omega(l)\nonumber\\ &&+c_3\bar \Gamma_1(l)\big[ \Lambda(l) +(l^2-\bar m^2-p^2) \Omega(l)\big] \Big\} -{k_f^3\over 3}(c_3+c_2\cos^2\psi)\sin^2 2\psi\bigg\}\,, \end{eqnarray}
with a suitable subtraction term to have only a logarithmic $\lambda$-dependence of the double-integral:
\begin{equation} {\pi k_f^3 \over 24 }\bigg[c_2\Big( 3m_\pi^2+{3k_f^2 \over 10}+{p^2 \over 2}+{q^2 \over 4}\Big)-4c_1 m_\pi^2+c_3\Big(6m_\pi^2+ {4k_f^2 \over 5}+{4p^2 +q^2 \over 3}\Big)  \bigg] \ln {m_\pi \over \lambda} \,. \end{equation} 
Under the same calculational treatment, the $c_4$-term in eq.(36) produces a further contribution to the isovector spin-orbit potential
\begin{equation} V_\text{med}^{(\text{cc})}= {c_4 \, \vec \tau_1\!\cdot\!\vec \tau_2\over 8\pi^5 f_\pi^6} i(\vec \sigma_1\!+\!\vec \sigma_2)\!\cdot\!(\vec q\!\times\!\vec p\,) \int_0^\lambda\!\! dr r \!\int_0^{\pi/2}\!\!\!d\psi \, { l_0^2 l \,\bar \Gamma_1(l)\over 4p^2-q^2} \Big[(p^2+l^2+\bar m^2)  \Omega(l)-\Lambda(l)\Big]  \,, \end{equation}
with a large-$\lambda$ behavior of the double-integral: $-(\pi k_f^3 /72) \ln(m_\pi/ \lambda)$. 
Before closing this section, let us have a closer look at the balance of  $\lambda^2$-divergences in the total sum $V_\text{med}^{(0)}+ V_\text{med}^{(1)} +V_\text{med}^{(\text{cc})}$. It reads:
\begin{equation} {k_f^3 \lambda^2 \over 192 \pi^4 f_\pi^6} \Big[ -  \vec \tau_1\!\cdot\!\vec \tau_2 (c_2+2c_3)+3\Big({c_2 \over 2}+c_3\Big) + \vec \tau_1\!\cdot\!\vec \tau_2(c_2+2c_3)\Big] = {k_f^3 \lambda^2 \over 128 \pi^4 f_\pi^6}(c_2+2c_3)\,, \end{equation} 
such that the remaining isoscalar piece can be absorbed on the 3N short-distance parameter 
$c_E$. This perfect matching gives an a posteriori justification to drop or subtract the $\lambda^2$-divergences at any place. In the case of the pieces proportional to $\ln (m_\pi /\lambda)$ one can verify  that these can be absorbed on the parameters $E_{1,\dots,10}$ of the subleading 3N-contact interaction \cite{3Ncontact} (see also eq.(49) in ref.\,\cite{vmedshort}).

\section{Ring interaction proportional to $g_A^2$}
The 3N-ring interaction proportional to $g_A^2c_{1,2,3,4}$ can be inferred from the coordinate-space potential written in eq.(4.7) of ref.\,\cite{midrange4}. By exploiting the permutational symmetry (and parity-invariance) one can obtain the following somewhat simpler form  
\begin{eqnarray}V_\text{3N}&=&-{g_A^2\over f_\pi^6}\int_0^\infty\!\! dl_0\!\int\!{ d^3l_2\over (2\pi)^4} {1\over (\bar m^2+l_1^2)(\bar m^2+l_2^2)(\bar m^2+l_3^2)} \Big\{\vec \tau_2\!\cdot\!\vec \tau_3\, \vec l_1\!\cdot\!(\vec l_2+\vec l_3)  \nonumber\\ && \times \big[2c_1 m_\pi^2+(c_2+c_3)l_0^2 +c_3 \,\vec l_2\!\cdot\!\vec l_3\big]+{c_4 \over 2}\Big[ \vec\tau_1\!\cdot\!(\vec \tau_2\!\times\!\vec \tau_3) \,\vec l_1\!\cdot\! \vec l_2 \,\vec \sigma_2\!\cdot\!(\vec l_1\!\times\!\vec l_3)+ \vec \tau_1\!\cdot\!(\vec\tau_2+\vec \tau_3)\nonumber\\ && \times \Big( \bar m^2(\vec \sigma_2\!\times\!\vec l_3)\!\cdot\!(\vec \sigma_3\!\times\!\vec l_2)  +\vec l_2\!\cdot\!\vec l_3\, \vec \sigma_2\!\cdot\!\vec l_1\, \vec \sigma_3\!\cdot\!\vec l_1 + \vec l_1\!\cdot\!\vec l_2\,\vec l_1\!\cdot\!\vec l_3 \, \vec \sigma_2\!\cdot\!\vec \sigma_3-2 \vec l_1\!\cdot\!\vec l_3\, \vec \sigma_2\!\cdot\!\vec l_2\, \vec \sigma_3\!\cdot\!\vec l_1  \Big) \Big]  \Big\}\,, \end{eqnarray}
which involves only three different isospin-operators: $\vec \tau_2\!\cdot\!\vec \tau_3,\, \vec\tau_1\!\cdot\!(\vec \tau_2\!\times\!\vec \tau_3)$ and $\vec \tau_1\!\cdot\!(\vec\tau_2+\vec \tau_3)$. Again, $\bar m$ stands for $\bar m = \sqrt{m_\pi^2+l_0^2}$ and one has to set $\vec l_1= \vec l_2-\vec q_3$ and $\vec l_3= \vec l_2+\vec q_1$.
 
\subsection{In-medium NN-potential from selfclosing of nucleon-lines}
Following the same procedure as in subsection\,3.1, one obtains from the selfclosing of nucleon-line\,1 (providing a non-vanishing spin-isospin trace) a further contribution to the isovector central potential:
\begin{eqnarray}V_\text{med}^{(0)} &=&{ g_A^2 k_f^3 \,\vec \tau_1\!\cdot\!\vec \tau_2 \over 48 \pi^4 f_\pi^6} \bigg\{ \Big[ 3m_\pi^2(4c_1-c_2-6c_3)-{5q^2 \over 6}(c_2 + 4c_3) \Big] \ln{m_\pi \over \lambda} +\Big(c_1+ {c_2 \over 12}+{11 c_3\over 6}\Big) m_\pi^2 \nonumber\\ && +\Big({41c_2 \over 8}\!+\!43 c_3 \Big) {q^2 \over 18} +\bigg[ {4m_\pi^2 \over 3}( 6c_1\!-\!c_2\!-\!7c_3) m_\pi^2 - {5 q^2 \over 6}(c_2 \!+\!4c_3) +{2m_\pi^2 q^2 \over 4m_\pi^2+q^2}(2c_1\!-\!c_3)\bigg] L(q)\bigg\}\,  \nonumber\\ \end{eqnarray}
where the $\lambda^2$-divergence has been dropped. Again, the appearance of the same coefficient $c_2+4c_3$ in the first and second line of eq.(52) guarantees the chiral limit $m_\pi\to 0 $ of $V_\text{med}^{(0)}$. 
\subsection{In-medium potential from concatenations $N_3$ on $N_2$ and $N_2$ on $N_3$ }
The concatenations $N_3$ on $N_2$ and $N_2$ on $N_3$ give for the first term with isospin-factor $\vec\tau_2\!\cdot\!\vec\tau_3$ in eq.(51) an isoscalar central potential of the form
\begin{eqnarray} V_\text{med}^{(1)}&=& {3 g_A^2\over 4\pi^5 f_\pi^6} \int_0^\lambda\!\! dr r \!\int_0^{\pi/2}\!\!\! d\psi \bigg\{l \bar \Gamma_1(l)\Big\{ {c_3 l \over 4p^2}( 4p^2-q^2) +\Big[ 2c_1 m_\pi^2+c_2 l_0^2 +c_3\Big( r^2-2 \bar m^2-p^2\nonumber\\ && +{q^2\over 4p^2}(l^2+\bar m^2-p^2)\Big) \Big] \Lambda(l) + \Big[ 2c_1 m_\pi^2+c_2 l_0^2 -c_3\Big(m_\pi^2+{q^2 \over 2}\Big) \Big] (l^2- \bar m^2-p^2) \Omega(l)\Big\}\nonumber\\ &&-{4 k_f^3\over 3}(c_3+c_2 \cos^2\psi) \sin^4\psi\bigg\}\,, \end{eqnarray}
where the subtraction term in the third line removes the $\lambda^2 k_f^3$-divergence. The remaining  logarithmic dependence of the double-integral on the cutoff $\lambda$ is
\begin{equation} {\pi k_f^3 \over 12 }\bigg[{c_2\over 4}\Big( 6m_\pi^2+k_f^2 +{5p^2 \over 3}+{q^2 \over 6}\Big) -6c_1 m_\pi^2 +c_3\Big(9m_\pi^2+k_f^2+{5p^2 \over 3}+q^2\Big) \bigg] \ln {m_\pi \over \lambda} \,. \end{equation}  
The second term $\sim c_4 \vec\tau_1\!\cdot\!(\vec\tau_2\!\times\!\vec\tau_3)$ in eq.(51) produces an isovector spin-orbit potential of the form
\begin{eqnarray} V_\text{med}^{(1)}&=& {c_4 g_A^2 \vec \tau_1\!\cdot\!\vec \tau_2\over 16\pi^5 f_\pi^6} i(\vec \sigma_1\!+\!\vec \sigma_2)\!\cdot\!(\vec q\!\times\!\vec p\,) \int_0^\lambda\!\! dr r \!\int_0^{\pi/2}\!\! \!d\psi \, { l\over 4p^2-q^2} \Big\{2 \bar \Gamma_2(l)\big[-2 \Lambda(l) \nonumber\\ &&+ (2l^2+2\bar m^2-2p^2+q^2)\Omega(l)\big] + \bar \Gamma_3(l)(\bar m^2+p^2-l^2) \big[ \Lambda(l)-(\bar m^2+l^2+p^2)\Omega(l)\big] \Big\}  \,, \end{eqnarray}
with a large-$\lambda$ behavior of the double-integral: $-(\pi k_f^3 /18) \ln(m_\pi/ \lambda)$.
The third term proportional to $c_4 \vec\tau_1\!\cdot\!(\vec\tau_2\!+\!\vec\tau_3)$ in eq.(51) gives on the one hand rise to an isovector central potential of the form
\begin{eqnarray} V_\text{med}^{(1)}&=& {c_4 g_A^2 \vec \tau_1\!\cdot\!\vec \tau_2\over 8\pi^5 f_\pi^6} \int_0^\lambda\!\! dr r \!\int_0^{\pi/2}\!\! \!d\psi \, \bigg\{2 l\big[\bar m^2 \bar \Gamma_0(l)+2 \bar \Gamma_2(l)\big] \big[2 \Lambda(l)- (2\bar m^2+q^2)\Omega(l)\big] \nonumber\\ &&+ l\Gamma_3(l)\Big[{l \over 2}+(3l^2-\bar m^2-p^2) \Lambda(l)+\Big( {B\over 2}-l^2(4\bar m^2+q^2)\Big)\Omega(l)\Big] -{8k_f^3 \over 3}
\sin^4\psi \bigg\}  \,, \end{eqnarray}
where the double-integral has the large-$\lambda$ behavior: 
$\pi k_f^3\big[ m_\pi^2+k_f^2/10 + p^2/6+7q^2/36\big] \ln(m_\pi/\lambda)$.
On the other hand one gets a contribution to the isovector spin-orbit potential of the form:
\begin{eqnarray} V_\text{med}^{(1)}&=& {c_4 g_A^2 \vec \tau_1\!\cdot\!\vec \tau_2\over 8\pi^5 f_\pi^6} i(\vec \sigma_1\!+\!\vec \sigma_2)\!\cdot\!(\vec q\!\times\!\vec p\,) \int_0^\lambda\!\! dr r \!\int_0^{\pi/2}\!\! \!d\psi \, { l\over 4p^2-q^2} \Big\{\big[\bar m^2 \bar \Gamma_0(l)+2 \bar \Gamma_2(l)\big]\big[2 \Lambda(l) \nonumber\\ &&+ (2p^2-2l^2-2\bar m^2-q^2)\Omega(l)\big] + \bar \Gamma_3(l)(\bar m^2+p^2-l^2) \big[(\bar m^2+l^2+p^2)\Omega(l)-\Lambda(l)\big] \Big\}  \,, \end{eqnarray}
with a large-$\lambda$ behavior  of the double-integral: $(\pi k_f^3 /24) \ln(m_\pi/ \lambda)$.
\subsection{In-medium NN-potential from remaining  four concatenations}
The other four concatenations, $N_3$ on $N_1$, $N_1$ on $N_3$, $N_1$ on $N_2$, and $N_2$ on $N_1$, applied the first term $\sim \vec\tau_2\!\cdot\!\vec\tau_3$ in eq.(51) produce an isovector central potential of the form
\begin{eqnarray} V_\text{med}^{(\text{cc})}&=& {g_A^2\vec\tau_1\!\cdot\!\vec \tau_2 \over 4\pi^5 f_\pi^6} 
\int_0^\lambda\!\! dr r \!\int_0^{\pi/2}\!\!\! d\psi \bigg\{ l  \big[ 2c_1 m_\pi^2+(c_2+c_3) l_0^2\big] 
\Big\{ \bar \Gamma_0(l)\big[2\Lambda(l)-(2\bar m^2+q^2)  \Omega(l)\big] \nonumber\\ &&+\bar \Gamma_1(l)\big[\Lambda(l)+(l^2-p^2-\bar m^2)  \Omega(l)\big]\Big\} +c_3 l \Big\{ \bar \Gamma_2(l)\big[ 2\Lambda(l)-(2
\bar m^2+q^2) \Omega(l)\big] \nonumber\\ &&+\bar \Gamma_3(l)\Big[{l \over 2} +(l^2-p^2-\bar m^2) \Lambda(l)+{1\over 2}(l^2-p^2-\bar m^2)^2 \Omega(l)\Big] +\bar \Gamma_1(l)\Big[{l \over 4p^2}(4p^2-q^2) \nonumber\\ &&+\Big( l^2-p^2-2\bar m^2-{q^2 \over 4} +{q^2 \over 4p^2}(l^2+\bar m^2) \Big) \Lambda(l)+\Big(\bar m^2+{q^2\over 2}\Big)(\bar m^2+p^2-l^2) \Omega(l)\Big] \Big\} \nonumber\\ && -{8k_f^3\over 3}(c_3+c_2\cos^2\psi)\sin^4\psi\bigg\}\,, \end{eqnarray}
where the double-integral depends (after subtraction in the last line) logarithmically on the cutoff: 
\begin{equation} {\pi k_f^3 \over 4 }\bigg[c_2\Big( m_\pi^2+{k_f^2 \over 10}+{p^2 \over 6}+{5q^2 \over 36}\Big)-4c_1 m_\pi^2 +c_3\Big(6m_\pi^2+ {11k_f^2 \over 15}+{11p^2 +5q^2 \over 9}\Big)  \bigg] \ln {m_\pi \over \lambda} \,, \end{equation} 
In the same way one obtains from the second term $\sim c_4 \vec\tau_1\!\cdot\!(\vec\tau_2\!\times\!\vec\tau_3)$ in eq.(51) a contribution to the isovector spin-orbit potential of the form
\begin{eqnarray} V_\text{med}^{(cc)}&=& {c_4 g_A^2 \vec \tau_1\!\cdot\!\vec \tau_2\over 8\pi^5 f_\pi^6} i(\vec \sigma_1\!+\!\vec \sigma_2)\!\cdot\!(\vec q\!\times\!\vec p\,) \int_0^\lambda\!\! dr r \!\int_0^{\pi/2}\!\! \!d\psi \, { l \,\bar \Gamma_1(l)\over 4p^2-q^2} \nonumber\\ && \times \Big\{(l^2+p^2) \Lambda(l)  -\big[(l^2-p^2)^2+q^2l^2+\bar m^2 (l^2+p^2)\big] \Omega(l)\Big\}  \,, \end{eqnarray}
with a large-$\lambda$ behavior  of the double-integral: $-(\pi k_f^3 /18) \ln(m_\pi/ \lambda)$. Under the same calculational treatment the third term $\sim c_4 \vec\tau_1\!\cdot\!(\vec\tau_2\!+\!\vec\tau_3)$ in 
eq.(51) gives rise to three spin-dependent potentials with the common isospin-factor $3+\vec \tau_1\!\cdot\!\vec \tau_2 $. The pertinent spin-spin potential has the form
\begin{eqnarray} V_\text{med}^{(cc)}&=& {c_4 g_A^2 \vec \sigma_1\!\cdot\!\vec \sigma_2\over 16\pi^5 f_\pi^6} (3+\vec \tau_1\!\cdot\!\vec \tau_2 )\int_0^\lambda\!\! dr r \!\int_0^{\pi/2}\!\! \!d\psi \bigg\{ l\bar\Gamma_1(l) \bigg[{l \over 2p^2}(3 p^2-l^2-\bar m^2 -q^2) \nonumber\\ && + \bigg(l^2-\bar m^2 -{3p^2 \over 2} +(\bar m^2 +l^2+q^2) {\bar m^2 +l^2\over 2p^2} + {q^2(4\bar m^2 +4l^2+q^2) \over 8p^2-2q^2} \bigg)\Lambda(l) \nonumber\\ && + {q^2(\bar m^2 +l^2+p^2) \over 4p^2-q^2}(2 p^2-2l^2-2\bar m^2 -q^2)  \Omega(l)\bigg] -{16 k_f^3 \over 9}\sin^4\psi \bigg\}  \,, \end{eqnarray}
with a  large-$\lambda$ behavior of the (subtracted) double-integral: $(\pi k_f^3 /9)\big[ 6m_\pi^2 +k_f^2+2p^2+3q^2/4\big] \ln(m_\pi/ \lambda)$. Note that the subtraction term acts only in the $^1\!S_0$ state with total isospin 1, therefore one can replace (of course only for this $\lambda^2 k_f^3$-term) the operator   $ \vec \sigma_1\!\cdot\!\vec \sigma_2(3+\vec \tau_1\!\cdot\!\vec \tau_2 )$ by $-3(3+\vec \tau_1\!\cdot\!\vec \tau_2 )$. Next, there is a 
tensor-type potential of the form
\begin{eqnarray} V_\text{med}^{(cc)}&=& {c_4 g_A^2 \over 8\pi^5 f_\pi^6} (3+\vec \tau_1\!\cdot\!\vec \tau_2 )(\vec \sigma_1\!\cdot\!\vec p\, \vec \sigma_2\!\cdot\!\vec p+\vec \sigma_1\!\cdot\!\vec p\,'\vec \sigma_2\!\cdot\!\vec p{\,'}) \int_0^\lambda\!\! dr r \!\int_0^{\pi/2}\!\! \!d\psi {l \bar\Gamma_1(l)\over 4p^2-q^2} \nonumber\\ && \times \bigg\{ {l \over 2p^2}\Big[3\bar m^2+3l^2\ -p^2+{5q^2\over 4}-{3q^2\over 4p^2}(\bar m^2+l^2)\Big]+ \bigg[{p^2\over 2}+l^2-\bar m^2-{5q^2\over 8}\nonumber\\ &&- {6(\bar m^2+l^2)^2+ q^2(\bar m^2+3l^2) \over 4p^2} +{3q^2\over 8p^4}(\bar m^2+l^2)^2- {q^2(4\bar m^2+4l^2+q^2)\over 4 p^2-q^2}\bigg]\Lambda(l) \nonumber\\ && +{q^2\over 4 p^2-q^2}\big[4(\bar m^2+l^2)^2+4 p^2(\bar m^2-l^2) + q^2(\bar m^2+3l^2+p^2)\big] \Omega(l)
\bigg\}  \,, \end{eqnarray}
with a large-$\lambda$ behavior of the double-integral: $-(\pi k_f^3 /36)\ln(m_\pi/ \lambda)$. Finally, one gets an ordinary tensor potential which has the form
\begin{eqnarray} V_\text{med}^{(cc)}&=& {c_4 g_A^2 \over 8\pi^5 f_\pi^6} (3+\vec \tau_1\!\cdot\!\vec \tau_2 )\vec \sigma_1\!\cdot\!\vec q\, \vec \sigma_2\!\cdot\!\vec q \int_0^\lambda\!\! dr r \!\int_0^{\pi/2}\!\! \!d\psi {l \bar\Gamma_1(l)\over 4p^2-q^2}\bigg\{{l\over 2p^2}(2p^2-q^2) \nonumber\\ &&+ \Big[{q^2(4\bar m^2 +4l^2+q^2)\over 8p^2-2q^2}+{\bar m^2+l^2\over 2p^2}(q^2-4p^2)\Big] \Lambda(l)+\Big[3(\bar m^2+l^2)^2 \nonumber\\ && +p^2(4\bar m^2+4l^2+p^2)+q^2(\bar m^2+2l^2+p^2)-{8 p^2\over 4 p^2-q^2}(\bar m^2+l^2+p^2)^2 \Big]  \Omega(l) \bigg\}  \,,  \end{eqnarray}
with a large-$\lambda$ behavior of the double-integral: $-(\pi k_f^3 /24)\ln(m_\pi/ \lambda)$.
Let us again consider the balance of $\lambda^2$-divergences in the total sum $V_\text{med}^{(0)}+ V_\text{med}^{(1)} +V_\text{med}^{(\text{cc})}$. With the equivalent form of the piece from eq.(61),  the balance reads
\begin{eqnarray} && {g_A^2k_f^3 \lambda^2 \over 96 \pi^4 f_\pi^6} \Big\{ -  \vec \tau_1\!\cdot\!\vec \tau_2 (c_2+6c_3)+\Big[ 3\Big({c_2 \over 2}+3c_3\Big) +3c_4 \vec \tau_1\!\cdot\!\vec \tau_2\Big]  +\Big[ \vec \tau_1\!\cdot\!\vec \tau_2(c_2+6c_3)-3c_4 (3+\vec\tau_1\!\cdot\!\vec \tau_2)\Big]\Big\} \nonumber \\&&  = {g_A^2 k_f^3 \lambda^2 \over 64 \pi^4 f_\pi^6}(c_2+6c_3-6c_4)\,, \end{eqnarray} 
and one observes that the remaining isoscalar piece can again be absorbed on the 3N short-distance parameter $c_E$. 
\section{Ring interaction proportional to $g_A^4$}
The 3N-ring interaction proportional to $g_A^4 c_{1,2,3,4}$ can be inferred from the coordinate-space potential written in eq.(4.6) of ref.\,\cite{midrange4}. In momentum-space this part of $V_\text{3N}$ at N$^4$LO is given by a euclidean loop-integral over three pion-propagators (one of them squared) times a long series of terms with different spin-, isospin- and momentum-dependence, which reads 
\begin{eqnarray}V_\text{3N}&=&{g_A^4 \over f_\pi^6}\int_0^\infty\!\! dl_0\!\int\!{ d^3l_2\over (2\pi)^4} {1\over (\bar m^2+l_1^2)(\bar m^2+l_2^2)^2(\bar m^2+l_3^2)} \nonumber\\ &&\times \Big\{\bar m^2\Big[ (\vec \sigma_1\!\times\!\vec l_3)\!\cdot \!(\vec \sigma_3\!\times\!\vec l_1)\big[ 2l_0^2(c_2+c_3)\vec \tau_1\!\cdot\!\vec \tau_3-6c_1 m_\pi^2 + \vec l_1\!\cdot\!\vec l_3 \big(c_4(\vec \tau_1\!+\!\vec \tau_3)\!\cdot\!\vec \tau_2- 3c_3\big)\big]\nonumber\\ && + 2 (\vec \sigma_1\!\times\!\vec l_2)\!\cdot \!(\vec \sigma_2\!\times\!\vec l_1)\big[ 2l_0^2(c_2+c_3) \vec \tau_1\!\cdot\!\vec \tau_2-6c_1 m_\pi^2 + \vec l_1\!\cdot\!\vec l_2 \big(c_4(\vec \tau_1\!+\!\vec \tau_2)\!\cdot\!\vec \tau_3- 3c_3\big)\big]\Big]\nonumber\\ && +{c_4\over 2} \vec l_1\!\cdot\!\vec l_2\big[ 2 \vec l_1\!\cdot\!\vec l_3 \,\vec \sigma_1\!\cdot\!(\vec l_3\!\times\!\vec l_2)-\vec l_2\!\cdot\!\vec l_3 \,\vec \sigma_2\!\cdot\!(\vec l_3\!\times\!\vec l_1)\big] \vec \tau_1\!\cdot\!(\vec \tau_2\!\times\!\vec \tau_3)+\vec l_1\!\cdot\!\vec l_2\,\vec l_1\!\cdot\!\vec l_3\, \vec l_2\!\cdot\!\vec l_3 \nonumber\\ && \times \big[ 2c_3 \vec \tau_1\!\cdot\!(2\vec \tau_2\!+\!\vec \tau_3) +2 \vec \sigma_2\!\cdot\!\vec \sigma_3\big( c_4 \vec \tau_1\!\cdot\!(\vec \tau_2\!+\!\vec \tau_3)-3c_3\big) +\vec \sigma_1\!\cdot\!\vec \sigma_3\big( c_4(\vec \tau_1\!+\!\vec \tau_3)\!\cdot\!\vec \tau_2 -3c_3\big)\big]  \nonumber\\ && +2 \vec l_1\!\cdot\!\vec l_2\,\vec l_1\!\cdot\!\vec l_3\big[ (c_2+c_3)l_0^2( 2 \vec \sigma_2\!\cdot\!\vec \sigma_3\,\vec \tau_2\!\cdot\!\vec \tau_3-3)-6c_1 m_\pi^2\vec \sigma_2\!\cdot\!\vec \sigma_3 +\vec \sigma_1\!\cdot\!\vec l_3\, \vec \sigma_2\!\cdot\!\vec l_2\big( 3c_3-c_4(\vec \tau_1\!+\!\vec \tau_2)\!\cdot\!\vec \tau_3\big) \big] \nonumber\\ && + \vec l_1\!\cdot\!\vec l_2\,\vec l_2\!\cdot\!\vec l_3\big[4c_1 m_\pi^2\vec \tau_1\!\cdot\!\vec \tau_3-3 l_0^2(c_2+ c_3)+ 2\vec \sigma_1\!\cdot\!\vec \sigma_3\big(l_0^2(c_2+ c_3)\vec \tau_1\!\cdot\!\vec \tau_3-3c_1 m_\pi^2\big)
+ 2\vec \sigma_1\!\cdot\!\vec l_1\,\vec \sigma_2\!\cdot\!\vec l_3 \nonumber\\ &&  \times \big( 3c_3-c_4(\vec \tau_1\!+\!\vec \tau_2)\!\cdot\!\vec \tau_3 \big) \big] + 2\vec l_1\!\cdot\!\vec l_3\,\vec l_2\!\cdot\!\vec l_3\big[4c_1 m_\pi^2\vec \tau_1\!\cdot\!\vec \tau_2 +\vec \sigma_1\!\cdot\!\vec l_1\,\vec \sigma_3\!\cdot\!\vec l_2\big(3c_3-c_4(\vec \tau_1\!+\!\vec \tau_3)\!\cdot\!\vec \tau_2 \big) \big] \nonumber\\ &&+(\vec l_1\!\cdot\!\vec l_3)^2 \vec \sigma_1\!\cdot\!\vec l_2\,\vec \sigma_3\!\cdot\!\vec l_2\big( c_4(\vec \tau_1\!+\!\vec \tau_3)\!\cdot\!\vec \tau_2 -3c_3\big)+ 2(\vec l_1\!\cdot\!\vec l_2)^2 \vec \sigma_1\!\cdot\!\vec l_3\,\vec \sigma_2\!\cdot\!\vec l_3\big( c_4(\vec \tau_1\!+\!\vec \tau_2)\!\cdot\!\vec \tau_3 -3c_3\big)\nonumber\\ &&+4\big( \vec l_2\!\cdot\!\vec l_3\, \vec \sigma_1\!\cdot\!\vec l_1\,\vec \sigma_2\!\cdot\!\vec l_3 +\vec l_1\!\cdot\!\vec l_3\, \vec \sigma_1\!\cdot\!\vec l_3\,\vec \sigma_2\!\cdot\!\vec l_2-  \vec l_1\!\cdot\!\vec l_2\, \vec \sigma_1\!\cdot\!\vec l_3\,\vec \sigma_2\!\cdot\!\vec l_3\big)\big( 3c_1m_\pi^2 - l_0^2 (c_2+c_3)  \vec \tau_1\!\cdot\!\vec \tau_2\big)\nonumber\\ &&+2\big( 2 \vec l_2\!\cdot\!\vec l_3\, \vec \sigma_1\!\cdot\!\vec l_1 -\vec l_1\!\cdot\!\vec l_3\, \vec \sigma_1\!\cdot\!\vec l_2\big)\vec \sigma_3\!\cdot\!\vec l_2 \big( 3c_1m_\pi^2 - l_0^2 (c_2+c_3) \vec \tau_1\!\cdot\!\vec \tau_3\big)\Big\}\,, \end{eqnarray}
with $\bar m = \sqrt{m_\pi^2+l_0^2}$ and one has to set $\vec l_1= \vec l_2-\vec q_3$ and $\vec l_3= \vec l_2+\vec q_1$. Without the prefactor $g_A^4/f_\pi^6$ the first line in eq.(65) defines a euclidean three-point function $\tilde J(q_1,q_2,q_3)$ that is symmetric under $q_1\leftrightarrow q_3$. By applying the Cutkosky cutting rule to the first and third pion-propagator, one can easily compute its imaginary part Im$\tilde J(q_1,\mu,q_3)$ as a $2\pi$-phase space integral over a squared pion-propagator, and obtains the following  spectral-representation 
\begin{equation}  \tilde J(q_1,q_2,q_3) = {1\over 16\pi^2}\!\int_{2m_\pi}^\infty \!\!\! d\mu\, { \mu^2 \sqrt{\mu^2-4m_\pi^2} \over \mu^2+q_2^2}\big[(\mu q_1q_3)^2+ m_\pi^2 G\big]^{-1}\,, \end{equation} 
 with the abbreviation $G =\big[\mu^2+(q_1+q_3)^2\big]\big[\mu^2+(q_1-q_3)^2\big]$. By a partial-fraction decomposition of the two denominators in eq.(66) one is able to find an analytical solution of the spectral-integral in terms of the even loop-function $L(s)=L(-s)$, defined in eq.(17). The final result for $\tilde J(q_1,q_2,q_3)$ reads:
\begin{equation} \tilde J(q_1,q_2,q_3) = {1\over 16\pi^2}\bigg\{{b_+^2 \big[L(q_2)-L(b_+)\big] \over (q_2^2-b_+^2)C} + {b_-^2\big[ L(b_-)-L(q_2)\big] \over (q_2^2-b_-^2)C}\bigg\} \,, \end{equation} 
with the auxiliary variables $b_\pm = \big( q_1 \sqrt{4m_\pi^2+q_3^2}\pm q_3 \sqrt{4m_\pi^2+q_1^2}\,\big)/(2m_\pi)$ and the symmetric combination $C = q_1 q_3 \sqrt{(4m_\pi^2+q_1^2)(4m_\pi^2+q_3^2)}$. By setting $q_j = 2m_\pi \sinh\zeta_j$ one is able to derive a representation of this three-point function entirely in terms of logarithms $L(q_j)$, which reads:
\begin{eqnarray} \tilde J(q_1,q_2,q_3) &\!\!=\!\!& {1\over 16\pi^2\Sigma}\bigg\{ \Big[2m_\pi^2+q_2^2-{2m_\pi^2\over 4m_\pi^2+q_1^2}(4m_\pi^2+q_2^2+q_3^2)\Big] L(q_1)\nonumber\\ && -q_2^2L(q_2)+ \Big[2m_\pi^2+q_2^2-{2m_\pi^2\over 4m_\pi^2 +q_3^2}(4m_\pi^2+q_1^2+q_2^2)\Big] L(q_3)\bigg\}\,, \nonumber  \end{eqnarray}
with $\Sigma= (q_1q_2q_3)^2+m_\pi^2\big[2q_1^2q_2^2+ 2q_1^2q_3^2+2q_2^2q_3^2 -q_1^4-q_2^4-q_3^4\big]>0$.
Likewise,  the (bare) euclidean loop-integral over three pion-propagators in the first line of eq.(51) defines a totally symmetric three-point function $J(q_1,q_2,q_3)$. It possesses a more involved spectral-representation 
\begin{equation} J(q_1,q_2,q_3) = {1\over 16\pi^2}\!\int_{2m_\pi}^\infty \!\!\! d\mu\, { \mu \over (\mu^2+q_2^2)\sqrt{G}} \ln{\mu(\mu^2 +q_1^2+q_3^2) +\sqrt{(\mu^2-4m_\pi^2)G}\over \mu(\mu^2 +q_1^2+q_3^2) - \sqrt{(\mu^2-4m_\pi^2)G}} \,, \end{equation} 
which does not allow for a solution in terms of elementary functions.
\subsection{In-medium NN-potential from selfclosing of nucleon-lines}
In this subsection the in-medium NN-potential $V_\text{med}^{(0)}$ is computed as it arises from the selfclosing of nucleon-lines for the (extremely lengthy) 3N-ring interaction written in eq.(65). One gets a non-vanishing spin-isospin trace from closing $N_1,\, N_2$ and $N_3$, respectively. After performing the angular and radial integrals the summed contributions are sorted according to their two-body spin- and isospin-operators. The complete list of  contributions to $V_\text{med}^{(0)}$ consists of an isoscalar central potential:
\begin{eqnarray}V_\text{med}^{(0)} &=&{ g_A^4(c_2+c_3) k_f^3\over 96 \pi^4 f_\pi^6} \bigg\{ -\bigg[ {135 
m_\pi^2\over 2} +{53q^2 \over 4} \bigg] \ln{m_\pi \over \lambda} -{73 m_\pi^2 \over 16} +{371 q^2 \over 96} \nonumber\\ && +\bigg[{12 m_\pi^4\over 4m_\pi^2+q^2} -35 m_\pi^2 -{53q^2 \over 4} \bigg] L(q)\bigg\}\,, \end{eqnarray}
an isovector central potential:
\begin{eqnarray}V_\text{med}^{(0)} &=&{ g_A^4 k_f^3\,\vec \tau_1\!\cdot\!\vec \tau_2 \over 96 \pi^4 f_\pi^6} \bigg\{ 15\Big[(7c_3-4c_1)m_\pi^2 +{23c_3 q^2 \over 18} \Big] \ln{m_\pi \over \lambda} +\Big( {389 c_3 \over 24}-27c_1\Big) m_\pi^2 +{8 m_\pi^4(2c_1-c_3)\over 4m_\pi^2+q^2}\nonumber\\ && -{553c_3 q^2 \over 144}+\bigg[4m_\pi^2\Big( {29 c_3 \over 3}-11c_1\Big) + {115c_3 q^2 \over 6} + {8 m_\pi^4(4c_1+3c_3)\over 4m_\pi^2 +q^2} + {32 m_\pi^6(2c_1-c_3)\over (4m_\pi^2 +q^2)^2} \bigg] L(q)\bigg\}\,, \nonumber\\ \end{eqnarray}
an isoscalar spin-spin potential: 
\begin{eqnarray}V_\text{med}^{(0)} &=&{ g_A^4 k_f^3\, \vec \sigma_1\!\cdot\!\vec \sigma_2 \over 96 \pi^4 f_\pi^6} \bigg\{ \Big[18(4c_1-5c_3)m_\pi^2 -11c_3 q^2\Big] \ln{m_\pi \over \lambda} +\Big( 18c_1-{35c_3 \over 2}\Big)m_\pi^2 +{c_3 q^2 \over 3}\nonumber\\ && +\bigg[ (48c_1-26c_3)m_\pi^2 -11c_3 q^2+{24m_\pi^4(c_3-2c_1)
\over 4m_\pi^2 +q^2}\bigg] L(q)\bigg\}\,,  \end{eqnarray}
an isovector  spin-spin potential:
\begin{eqnarray}V_\text{med}^{(0)} &=&{ g_A^4(c_2+c_3) k_f^3\over 48 \pi^4 f_\pi^6} \vec \sigma_1\!\cdot\!\vec \sigma_2\,\vec \tau_1\!\cdot\!\vec \tau_2 \, \bigg\{ (6m_\pi^2 +q^2) \ln{m_\pi \over \lambda} +{m_\pi^2 \over 2} -{q^2 \over 6}+(      2m_\pi^2 +q^2) L(q)\bigg\}\,, \end{eqnarray}
an isoscalar tensor potential:
\begin{eqnarray}V_\text{med}^{(0)} &=&{ g_A^4 k_f^3\over 48 \pi^4 f_\pi^6} \vec \sigma_1\!\cdot\!\vec q\, \vec \sigma_2\!\cdot\!\vec q  \, \bigg\{ 4c_3 \ln{m_\pi \over \lambda} +{61 c_3 \over 48} +{2m_\pi^2 \over q^2}(3c_1+c_3) \nonumber\\ && + \bigg[ 4c_3 -{2m_\pi^2 \over q^2}(3c_1+c_3) +{3m_\pi^2 (c_3-2c_1)\over 4m_\pi^2+q^2}\bigg]  L(q)\bigg\}\,, \end{eqnarray}
and an isovector tensor potential:
\begin{equation}V_\text{med}^{(0)} ={ g_A^4(c_2+c_3) k_f^3\over 72 \pi^4 f_\pi^6} \vec \sigma_1\!\cdot\!\vec q \, \vec \sigma_2\!\cdot\!\vec q \,\,\vec \tau_1\!\cdot\!\vec \tau_2 \,\bigg\{ - \ln{m_\pi \over \lambda}-{7 \over 48} +{m_\pi^2 \over q^2} -\Big(1+{m_\pi^2 \over q^2}\Big) L(q)\bigg\}\,. \end{equation}
Note that $c_4$ has dropped out and the dependence on the other three low-energy constants $c_{1,2,3}$ is well structured. The isoscalar central and isovector spin-dependent potentials are solely proportional to the sum $c_2+c_3$, whereas the other potentials depend separately on $c_1$ and $c_3$. The total $\lambda^2$-divergence behind the central and spin-spin potentials written in eqs.(69-72) is (for S-waves, where the replacements $\vec\sigma_1\!\cdot\!\vec\sigma_2\, \vec\tau_1\!\cdot\!\vec\tau_2 \to -3$ and $\vec\sigma_1\!\cdot\!\vec\sigma_2 \to -2-\vec\tau_1\!\cdot\!\vec\tau_2$ apply) equivalent to \begin{equation}{g_A^4 k_f^3 \lambda^2 \over (4\pi)^4 f_\pi^6} \bigg\{ {83 c_3 \over 2} -{77 c_2 \over 2}+ 75 c_3 \,\vec \tau_1\!\cdot\!\vec \tau_2\bigg\}\,. \end{equation}
In combination with the $\lambda^2$-divergences from concatenations $V_\text{med}^{(2)}+V_\text{med}^{(cc)}$ for all interaction terms in eq.(65) (several examples are given in the next subsection) it reduces to an isoscalar component only:
\begin{eqnarray}&& {g_A^4 k_f^3 \lambda^2 \over (4\pi)^4 f_\pi^6} \bigg\{ {83 c_3 \over 2} -{77 c_2 \over 2}+ 75 c_3 \,\vec \tau_1\!\cdot\!\vec \tau_2+(1+2)\bigg[{13c_2\over 12}-{277c_3\over 12}+40c_4-25c_3 \,\vec \tau_1\!\cdot\!\vec \tau_2\bigg]\bigg\} \nonumber \\ && = {3g_A^4 k_f^3 \lambda^2 \over (4\pi)^4 f_\pi^6}\Big[40c_4-{1\over 4}(47c_2+37c_3)\Big]\,, \end{eqnarray}
that can be absorbed on the 3N short-distance parameter $c_E$. This property
serves as a good check on our calculation.
\subsection{In-medium NN-potential from concatenations for three selected terms}
The 3N-ring interaction  $V_\text{3N}$ written in eq.(65) consists of a very large number of terms. In this paper we consider  for the contributions to $V_\text{med}$ from concatenations of two nucleon-lines only for three selected terms. The analogous formulas for all the other terms can be obtained from the author upon request.\\

\noindent
a) The term proportional to $c_4 \vec \tau_1\!\cdot\!(\vec \tau_2\!\times\!\vec\tau_3)$ in the fourth line of eq.(65). It gives for the concatenations $N_3$ on $N_1$ and  $N_1$ on $N_3$ an isovector spin-orbit potential of the form
\begin{eqnarray} V_\text{med}^{(2)}&=& {c_4 g_A^4 \vec \tau_1\!\cdot\!\vec \tau_2\over 16\pi^5 f_\pi^6} i(\vec \sigma_1\!+\!\vec \sigma_2)\!\cdot\!(\vec q\!\times\!\vec p\,) \int_0^\lambda\!\! dr r \!\int_0^{\pi/2}\!\! \!d\psi \, { l \over 4p^2-q^2} \bigg\{ \bar \gamma_2(l)\bigg[ {l\over p^2}(4p^2-q^2)  + \bigg(4p^2 \nonumber\\ && +{q^2\over p^2}(\bar m^2+l^2) -4l^2-8\bar m^2-3q^2\bigg)\Lambda(l) +(2\bar m^2 +q^2)(2\bar m^2+2l^2 -2p^2+q^2) \Omega(l)\bigg]\nonumber\\ && + \bar \gamma_3(l)\bigg[{l\over 4p^2}(4p^2-q^2)(\bar m^2+ l^2-p^2)+ \bigg({q^2\over 4p^2}(\bar m^2+l^2)^2+2(p^2-l^2)(\bar m^2+l^2+p^2)\nonumber\\ && +{q^2\over 4}(2\bar m^2-2l^2+ p^2)\bigg)\Lambda(l) +(l^2-\bar m^2-p^2)\Big( B +{q^2 \over 2}(\bar m^2+3l^2+p^2)\Big) \Omega(l)\bigg] \bigg\}  \,, \end{eqnarray}
with a large-$\lambda$ behavior  of the double-integral: $(5\pi k_f^3 /72) \ln(m_\pi/ \lambda)$. The new functions $\bar\gamma_{2,3}(l)$ appearing in eq.(76) are $\bar\gamma_{2,3}(l)=-\partial\bar\Gamma_{2,3}(l)/\partial\bar m^2$ with $\bar\Gamma_{2,3}(l)$ given in eqs.(40,41). The other four concatenations produce also an isovector spin-orbit potential of the form:
\begin{eqnarray} V^\text{(cc)}_\text{med} &=& {c_4g_A^4 \vec\tau_1\!\cdot\!\vec \tau_2\over  16\pi^5 f_\pi^6} i (\vec\sigma_1\!+\!\vec\sigma_2)\!\cdot\!(\vec q \!\times \! \vec p\,)\int_0^\lambda\!\! dr r \!\int_0^{\pi/2}\!\! \!d\psi \,\bigg\{l \bar \Gamma_2(l)\bigg\{\bigg[{1\over p^2}+{4 \over 4p^2-q^2}\bigg] \Lambda(l)- {l\over B}\bigg[ {\bar m^2+l^2\over p^2}\nonumber \\ &&-1+{2\bar m^2+q^2\over B+q^2l^2} (\bar m^2+p^2-l^2)\bigg] +\bigg[2-{4(\bar m^2 +l^2+p^2)\over 4p^2-q^2} +{2\bar m^2+q^2\over B+q^2l^2}(l^2-\bar m^2-p^2)\bigg] \Omega(l) \bigg\} \nonumber \\ && +{l \over 2}\bar\Gamma_3(l)\bigg\{ {l(\bar m^2+p^2-l^2)\over B+q^2l^2} -{l \over p^2} +\bigg( {\bar m^2+l^2-p^2\over p^2} -{4\bar m^2+2q^2\over 4p^2-q^2}\bigg) \Lambda(l)\nonumber \\ &&+
\bigg[ {\bar m^2+l^2+p^2\over 4p^2-q^2}(4\bar m^2+8p^2)-\bar m^2-3l^2-p^2+{q^2l^2(l^2-\bar m^2-p^2) \over B+q^2l^2 }\bigg] \Omega(l) \bigg\} \bigg\}\,,\end{eqnarray}
with a large-$\lambda$ behavior  of the double-integral: $(5\pi k_f^3 /36) \ln(m_\pi/ \lambda)$.\\

\noindent
b) The term proportional to $\vec l_1\!\cdot\!\vec l_2\,\vec l_1\!\cdot\!\vec l_3\, \vec l_2\!\cdot\!\vec l_3$ multiplied by the fifth line in eq.(65). It gives for the concatenations $N_3$ on $N_1$ and $N_3$ on $N_1$ a combination of central and spin-spin potentials of the form 
\begin{eqnarray} V^\text{(2)}_\text{med} &=& {g_A^4 \over  8\pi^5 f_\pi^6}\Big[ {3c_3\over 2}+3(c_3-c_4) \vec\sigma_1\!\cdot\!\vec\sigma_2 -(2c_3+3c_4) \vec\tau_1\!\cdot\!\vec \tau_2 -c_4 \vec\sigma_1\!\cdot \!\vec\sigma_2\, \vec\tau_1\!\cdot\!\vec \tau_2\Big]  \int_0^\lambda\!\! dr r \!\int_0^{\pi/2}\!\! \!d\psi \nonumber \\ && \times\bigg\{l \bar \gamma_2(l) \bigg[{l\over p^2}(4p^2-q^2)+\Big( {q^2\over p^2}(\bar m^2 +l^2)-8\bar m^2-3q^2\Big)  \Lambda(l)+(2 \bar m^2 + q^2)^2\, \Omega(l) \bigg] \nonumber \\ && +l\bar\gamma_3(l)\bigg[ l(3l^2-p^2-2\bar m^2)+{l q^2\over 2p^2}(\bar m^2-l^2) +\Big(p^2-l^2+3\bar m^2+{q^2 \over 2} -{q^2 \over 2p^2}(\bar m^2+l^2)\Big) \nonumber \\ && \times (\bar m^2+p^2-l^2) \Lambda(l) - \Big(\bar m^2+{q^2\over 2}\Big)(\bar m^2+p^2-l^2)^2\,  \Omega(l) \bigg] -{8k_f^3 \over 3} \sin^8\psi \bigg\}\,,\end{eqnarray}
with a large-$\lambda$ behavior of the double-integral: $(\pi k_f^3 /24) \big[ 35m_\pi^2+18k_f^2/5+6p^2+43q^2/12\big]  \ln(m_\pi/ \lambda)$. The other four concatenations produce the same  combination of central and spin-spin potentials, which takes the form:
\begin{eqnarray} V^\text{(cc)}_\text{med} &=& {g_A^4 \over  8\pi^5 f_\pi^6}\Big[ {3c_3\over 2}+3(c_3-c_4) \vec\sigma_1\!\cdot\!\vec\sigma_2 -(2c_3+3c_4) \vec\tau_1\!\cdot\!\vec \tau_2 -c_4 \vec\sigma_1\!\cdot \!\vec\sigma_2\, \vec\tau_1\!\cdot\!\vec \tau_2\Big]  \int_0^\lambda\!\! dr r \!\int_0^{\pi/2}\!\! \!d\psi \nonumber \\ && \times\bigg\{l \bar \Gamma_2(l) \bigg[{l\over B}\bigg( {q^2\over p^2}(\bar m^2 +l^2)-8\bar m^2-3q^2+{ \bar m^2 +l^2+p^2 \over B+q^2l^2}(2\bar m^2 +q^2)^2 \bigg)\nonumber \\ &&+\Big(8-{q^2\over p^2}\Big)\Lambda(l)+\bigg( {2 \bar m^2 + q^2 \over B+q^2l^2} (\bar m^2 +l^2+p^2)-4\bigg) (2 \bar m^2 + q^2) \Omega(l) \bigg] \nonumber \\ && +l\bar\Gamma_3(l)\bigg[{l \over B}\bigg(3(p^2-l^2)^2+\bar m^2(4 \bar m^2-l^2+7p^2) +{l^2(2\bar m^2+q^2)\over B+q^2l^2}(\bar m^2+l^2+p^2)\Big(2\bar m^2+ {q^2\over 2}\Big)\nonumber \\ &&-{\bar m^2 q^2\over p^2}(\bar m^2+l^2) -{q^2\over 2}(3\bar m^2+l^2+p^2)\bigg) +\Big(4l^2-4p^2-6\bar m^2+{\bar m^2q^2 \over p^2} \Big) \Lambda(l) +\bigg(3\bar m^2-l^2\nonumber \\ && +p^2+q^2+{2\bar m^2 + q^2 \over 2(B+q^2l^2)}\big(l^4-(\bar m^2+p^2)^2\big)\bigg)(\bar m^2+p^2-l^2)\Omega(l) \bigg] 
-{16k_f^3 \over 3} \sin^8\psi \bigg\}\,,\end{eqnarray}
with a large-$\lambda$ behavior  of the double-integral: $(\pi k_f^3 /12) \big[ 35m_\pi^2+79k_f^2/20+79p^2/12+3q^2\big] \ln(m_\pi/ \lambda)$. 

\noindent
c) The spin- and isospin-dependent term in the last line of eq.(65). It gives
for the concatenations $N_3$ on $N_1$ and $N_3$ on $N_1$ an isoscalar central
potential:
\begin{eqnarray} V^\text{(2)}_\text{med} &=& {3g_A^4 \over  4\pi^5 f_\pi^6} \int_0^\lambda\!\! dr r \!\int_0^{\pi/2}\!\! \!d\psi\,\bigg\{ \big[(c_2+c_3)l_0^2-c_1
m_\pi^2\big]\bigg[ l \bar \gamma_2(l)\Big((2\bar m^2+q^2)\Omega(l)-2\Lambda(l)   \Big) +  l \bar \gamma_3(l)\nonumber \\ && \times \Big(l - 2(\bar m^2+p^2) \Lambda(l)+\big(l^2(q^2-2\bar m^2)+B\big)  \Omega(l)\Big) \bigg] -{4k_f^3 \over 3}(c_2+c_3)\sin^6\psi\cos^2\psi\bigg\}\,, \end{eqnarray}
with a large-$\lambda$ behavior of the double integral:
\begin{equation} {\pi k_f^3 \over 48}\bigg[ 10c_1 m_\pi^2+(c_2+c_3)\bigg(5m_\pi^2+{9 k_f^2 \over 5}+3p^2-{7q^2 \over 12}\bigg) \bigg] \ln{m_\pi \over \lambda}\,,
\end{equation}
and an isoscalar spin-orbit potential:
\begin{eqnarray} V_\text{med}^{(2)}&=& {3g_A^4 \over 2\pi^5 f_\pi^6} i(\vec \sigma_1\!+\!\vec \sigma_2)\!\cdot\!(\vec q\!\times\!\vec p\,) \int_0^\lambda\!\! dr r \!\int_0^{\pi/2}\!\! \!d\psi \,\big[(c_2+c_3)l_0^2-c_1m_\pi^2\big] { l \over 4p^2-q^2} \bigg\{ \bar \gamma_2(l) \Big[\big(2\bar m^2 +q^2\nonumber \\ &&+2l^2-2p^2\big)\Omega(l)-2 \Lambda(l) \Big]+ {\bar \gamma_2(l)\over 2}(\bar m^2+p^2-l^2)\Big[ \Lambda(l)-(\bar m^2+p^2+l^2)\Omega(l)\Big] \bigg\}\,,\end{eqnarray}
with a large-$\lambda$ behavior of the double integral: $-(\pi k_f^3/576)(c_2+c_3)\ln(m_\pi/\lambda)$. The other four concatenations produce a spin-spin
potential:
\begin{eqnarray} V^\text{(cc)}_\text{med} &=& {g_A^4 \vec \sigma_1\!\cdot\!\vec \sigma_2\over  2\pi^5 f_\pi^6} \int_0^\lambda\!\! dr r \!\int_0^{\pi/2}\!\! \!d\psi
\,\bigg\{ \big[(c_2+c_3)l_0^2 \vec \tau_1\!\cdot\!\vec \tau_2-3c_1m_\pi^2\big]  {l \bar \Gamma_1(l)\over 4p^2-q^2}\bigg[ \Big(3\bar m^2+2l^2+2p^2+{q^2\over 2}\Big)\Lambda(l)\nonumber \\ &&- \bigg(2B+(\bar m^2+p^2+l^2)\Big(\bar m^2+{q^2 \over 2}\Big)+  2q^2l^2\bigg) \Omega(l)\bigg]-{4k_f^3 \over 9}(c_2+c_3)\vec \tau_1\!\cdot\!\vec \tau_2\sin^6\psi\cos^2\psi\bigg\}\,, \end{eqnarray}
with a large-$\lambda$ behavior of the double integral:
\begin{equation} {\pi k_f^3 \over 24}\bigg[ 5c_1m_\pi^2+ (c_2+c_3)\vec \tau_1\!\cdot\!\vec \tau_2\bigg( {5m_\pi^2 \over 6}+{11\over 120}(k_f^2+p^2)+{q^2\over 5}\bigg)\bigg] \ln{m_\pi\over \lambda}\,, \end{equation}
an ordinary tensor potential:
\begin{eqnarray} V^\text{(cc)}_\text{med} &=& {g_A^4 \over  2\pi^5 f_\pi^6} \vec  \sigma_1\!\cdot\!\vec q\, \vec \sigma_2\!\cdot\!\vec q\int_0^\lambda\!\! dr r \! \int_0^{\pi/2}\!\! \!d\psi\,\big[(c_2+c_3)l_0^2 \vec \tau_1\!\cdot\!\vec \tau_2-  3c_1m_\pi^2\big]  {l \bar \Gamma_1(l)\over 4p^2-q^2}\bigg\{{l \over B}\nonumber \\ && \times  \bigg[ {2p^2-\bar m^2-q^2\over B+q^2l^2} \Big((\bar m^2+p^2)^2+l^2\Big(\bar m^2-p^2+{q^2 \over 2}\Big) \Big)+{\bar m^2+q^2\over 2}-l^2
 \nonumber \\ &&  -{1\over q^2}\Big(2l^4 +l^2(5\bar m^2-6p^2)+3\bar m^4+5\bar m^2p^2+4p^4\Big)\bigg]+\bigg[ {6\bar m^2+4l^2+2q^2\over  4p^2-q^2}-{1\over 2}\bigg] \Lambda(l) \nonumber \\ &&+\bigg[ {2p^2-\bar m^2-q^2\over B+q^2l^2}\Big((\bar m^2+p^2)^2+   l^2\Big(\bar m^2-p^2+{q^2 \over 2}\Big) \Big)-{2(\bar m^2+p^2+l^2)\over 4p^2-q^2}(3\bar m^2+2l^2+q^2)\nonumber \\ &&+{1\over q^2}\Big(2l^4+l^2(5\bar m^2\!-\!6p^2) +3\bar m^4+5\bar m^2p^2+4p^4\Big)+2\bar m^2+{1\over 2}(11l^2\!-\!7p^2\!+\!3q^2) \bigg]     \Omega(l) \bigg\}\,,\end{eqnarray}
with a large-$\lambda$ behavior of the double integral: $(\pi k_f^3/160)(c_2+
c_3)\vec \tau_1\!\cdot\!\vec \tau_2\ln(m_\pi/\lambda)$,  and a tensor-type potential:
\begin{eqnarray} V^\text{(cc)}_\text{med} &=& {g_A^4 \over  2\pi^5 f_\pi^6} (\vec  \sigma_1\!\cdot\!\vec p\, \vec \sigma_2\!\cdot\!\vec p+\vec  \sigma_1\!\cdot\!\vec p\,' \vec \sigma_2\!\cdot\!\vec p\,')\int_0^\lambda\!\! dr r \! \int_0^{\pi/2}\!\! \!d\psi\,\big[(c_2+c_3)l_0^2 \vec \tau_1\!\cdot\!\vec \tau_2-  3c_1m_\pi^2\big]  {l \bar \Gamma_1(l)\over 4p^2-q^2}\bigg\{{l \over B}  \nonumber \\ && \times \bigg[{2\bar m^2+q^2\over B+q^2l^2} \Big((\bar m^2+p^2)^2+l^2\Big(\bar m^2-p^2+{q^2 \over 2}\Big) \Big)+{3\bar m^2-q^2 \over 2}-l^2+{1\over 2p^2}(2l^4+5l^2\bar m^2+3\bar m^4)\bigg] \nonumber \\ &&-\bigg[ {4\over  4p^2-q^2}(3\bar m^2+2l^2+q^2 )+1 +{3\bar m^2+2l^2\over 2p^2}\bigg] \Lambda(l)  +\bigg[{4(\bar m^2+p^2+l^2)\over  4p^2-q^2}(3\bar m^2+2l^2+q^2) \nonumber \\ && +{2\bar m^2+q^2\over B+q^2l^2} \Big((\bar m^2+p^2)^2+l^2\Big(\bar m^2-p^2+{q^2 \over 2}\Big) \Big) +p^2-5l^2-3\bar m^2-{3q^2\over 2}\bigg]\Omega(l) \bigg\}\,,\end{eqnarray}
with a large-$\lambda$ behavior of the double integral: $(11\pi k_f^3/2880)(c_2+
c_3)\vec \tau_1\!\cdot\!\vec \tau_2\ln(m_\pi/\lambda)$.
\section{Summary and outlook}
In this work the density-dependent in-medium NN-interaction $V_\text{med}$ has
been derived from the subsubleading chiral 3N-forces. This is necessary since
for the intermediate-range topologies ($2\pi1\pi$-exchange and ring-diagrams)
the N$^4$LO corrections of ref.\,\cite{midrange4} dominate in most cases over the
nominally leading N$^3$LO terms. The loop-integrals representing the 3N-ring
interaction proportional to $c_{1,2,3,4}$ have been regularized by a (euclidean)
cutoff $\lambda$ and each contribution to $V_\text{med}$ has been presented such
that the absorption of ($\lambda^2$  and $\ln(m_\pi/\lambda)$) divergences
on the 3N short-distance parameters becomes obvious. In the next step,
partial-wave matrix elements of $V_\text{med}$ will be calculated numerically
\cite{Vmedreview} in order to study quantitatively the effects of the
subleading \cite{vmedshort,vmedlong} as well as subsubleading chiral 3N-forces.
At the same time the construction of 3N-forces in chiral effective field theory
with explicit $\Delta(1232)$-isobars by the Bochum group should be accompanied
by a calculation of the corresponding density-dependent NN-potential $V_\text{med}$. On the other hand,
neutron matter calculations with (sub)-subleading
chiral 3n-forces require the $\rho_n$-dependent
nn-interaction in pure neutron matter.

\section*{Acknowledgements}
I thank  H. Krebs for detailed information on the chiral three-nucleon forces
at N$^4$LO.

\end{document}